\documentclass[useAMS,usenatbib]{mn2e}

\usepackage{graphics}
\usepackage{graphicx}
\usepackage{listings}
\usepackage{amssymb}
\usepackage{amsmath}
\title[A Keplerian-based Splitting for $N$-body Simulations]{A Keplerian-based Hamiltonian Splitting for Gravitational $N$-body Simulations}

\author[G. Gon\c calves Ferrari et al.]{G. Gon\c calves Ferrari$^{1,2}$\thanks{E-mail: gg.ferrari@gmail.com (GGF); spz@strw.leidenuniv.nl (SPZ); boekholt@strw.leidenuniv.nl (TB)}, T. Boekholt$^{1}$\footnotemark[1] and S.F. Portegies Zwart$^{1}$\footnotemark[1]\\
$^{1}$Leiden Observatory, Leiden University, P.O. Box 9513, 2300 RA Leiden, The Netherlands\\
$^{2}$Instituto de F\'{\i}sica, Universidade Federal do Rio Grande do Sul, Porto Alegre, Brazil.}

\begin{document}

\date{Accepted 2014 February 11. Received 2014 February 5; in original form 2013 February 17}

\pagerange{\pageref{firstpage}--\pageref{lastpage}} \pubyear{2014}

\maketitle

\label{firstpage}

\begin{abstract}
We developed a Keplerian-based Hamiltonian splitting for solving the gravitational $N$-body problem. This splitting allows us to approximate the solution of a general $N$-body problem by a composition of multiple, independently evolved $2$-body problems. While the Hamiltonian splitting is exact, we show that the composition of independent $2$-body problems results in a non-symplectic non-time-symmetric first-order map. A time-symmetric second-order map is then constructed by composing this basic first-order map with its self-adjoint. The resulting method is precise for each individual $2$-body solution and produces quick and accurate results for near-Keplerian $N$-body systems, like planetary systems or a cluster of stars that orbit a supermassive black hole. The method is also suitable for integration of $N$-body systems with intrinsic hierarchies, like a star cluster with primordial binaries. The superposition of Kepler solutions for each pair of particles makes the method excellently suited for parallel computing; we achieve $\gtrsim 64\%$ efficiency for only $8$ particles per core, but close to perfect scaling for $16384$ particles on a $128$ core distributed-memory computer. We present several implementations in \texttt{Sakura}, one of which is publicly available via the AMUSE framework.
\end{abstract}


\section{Introduction}
\label{sec1:introduction}

Since the pioneering work of \cite{vonHoerner60}, \cite{Aarseth63} and \cite{vanAlbada68} $N$-body simulations have been an essential tool for the theoretical understanding of self-gravitating astrophysical systems. Such systems often show a large dynamic range of time-scales. Thus, instead of a fixed or adaptive global time-step, most of the $N$-body codes adopt individual or block time-step algorithms in order to advance the particles in time \citep{McMillan86, MakinoAarseth92, Aarseth03}. In addition, different approaches to calculate the acceleration of each particle, such as using grids \citep{HockneyEastwood88} or a hierarchical tree data structure \citep{BarnesHut86}, are commonly employed to decrease the computational cost of the simulations. These approaches allow the use of a larger number of particles, despite only giving an approximation to the true acceleration of each particle. Therefore, these codes should not inadvertently be used in simulations of collisional systems such as planetary systems, dense star clusters or the inner parts of galactic nuclei.

In collisional systems the individual interactions between particles play an important role in the dynamical evolution of the system as a whole. For example, the formation of hard binaries in star cluster core collapse \citep{Tanikawa12} requires very precise integration methods to correctly evolve close encounters between particles. This precision is only possible if we use more accurate, direct brute-force methods, to calculate the accelerations due to each pair of particles in the system. The main difficulty here is that with the formation of the first hard binary in the system, the simulation as a whole experiences a slow-down in performance due to the necessity to decrease the time-step size in order to accurately integrate such compact sub-systems.

Currently, the  most effective and common approach to overcome such obstacles seems to be a combination of the block time-step algorithm \citep{McMillan86}, Ahmad-Cohen neighbour scheme \citep{AhmadCohen73} and some sort of $2$-body regularization \citep{PretoTremaine1999, MikkolaTanikawa1999a, MikkolaTanikawa1999b, MikkolaAarseth2002} in order to handle very compact sub-systems efficiently. This is the approach used in modern Hermite integrators for collisional stellar systems \citep{MakinoAarseth92, Aarseth03}.

In this paper, we develop a new Keplerian-based Hamiltonian splitting for the gravitational $N$-body problem. This splitting allows us to approximate the solution of a general $N$-body problem by a composition of independently evolved $2$-body problems. While the Hamiltonian splitting is exact, we show in section~\ref{sec2:method} that the composition of independent $2$-body problems results in a non-symplectic non-time-symmetric first-order map. A time-symmetric second-order map is then constructed by composing this basic first-order map with its self-adjoint. The advantages of this Keplerian-based integrator are: i) a guarantee that every pair of particles is always integrated precisely; ii) the method does not suffer from slow-down in performance when tight binaries are present in the simulation, and iii) the method allows for good parallel efficiency \citep{PortegiesZwart2001}.

\section{Method}
\label{sec2:method}

\subsection{Hamiltonian splitting}
\label{sec2.1:splitting}

We begin the derivation of our scheme for the numerical integration of a gravitational $N$-body system by considering its Hamiltonian,
\begin{eqnarray}
H = H_{T} + H_{U}\,.\label{eq:01}
\end{eqnarray}
Here,
\begin{eqnarray}
H_{T} \equiv \sum_{i=1}^{N} H_{T_{i}}\,,\qquad H_{T_{i}} \equiv \frac{1}{2}m_{i}v^{2}_{i}\,,\label{eq:02}
\end{eqnarray}
and
\begin{eqnarray}
H_{U} \equiv \frac{1}{2}\sum_{i=1}^{N}\sum_{j\ne i}^{N} H_{U_{ij}}\,,\qquad H_{U_{ij}} \equiv -\frac{m_{i} m_{j}}{r_{ij}}\,,\label{eq:03}
\end{eqnarray}
are the kinetic and potential energies of the system, respectively; $m_{i}$ and $v_{i}=|\mathbf{v}_{i}|$ are the mass and velocity of the $i$-th particle and $r_{ij}=|\mathbf{r}_{ij}|=|\mathbf{r}_{i}-\mathbf{r}_{j}|$ is the relative distance between particles $i$ and $j$.

The time evolution of a Hamiltonian system is formally given by the operator\footnote{Hamiltonian associated operators are denoted by a $\widehat{\ }$ symbol.} $e^{\tau \widehat{H}}$, which can be approximated by a composition of individually solvable operators $e^{\tau \widehat{H}_{A}}$ and $e^{\tau \widehat{H}_{B}}$ in cases when the Hamiltonian can be split as $\widehat{H} = \widehat{H}_{A} + \widehat{H}_{B}$. The simplest example of Hamiltonian splitting is the case when $\widehat{H}_{A} = \widehat{H}_{T}$ and $\widehat{H}_{B} = \widehat{H}_{U}$, for which we can generate the time-symmetric second-order Drift-Kick-Drift (DKD) variant of the Leapfrog integrator: $e^{\tau \widehat{H}} \approx e^{\frac{\tau}{2} \widehat{H}_{T}} e^{\tau \widehat{H}_{U}} e^{\frac{\tau}{2} \widehat{H}_{T}}$. This Hamiltonian splitting is not the only possibility and many other ways of subdividing the system have been tried \citep{WisdomHolman91, Duncan_etal98, Chambers99, Fujii_etal07, Pelupessy_etal12}.

In the present paper we introduce a way to split the Hamiltonian of an $N$-body system, which is based on two main arguments: i) the validity of the superposition principle\footnote{Recall that the gravitational potential and acceleration at the position of a given particle consists of a superposition of $2$-body contributions due to the interaction with every other particle in the system.}, and ii) the existence of an analytical solution for the $2$-body problem. Therefore, a natural way to approximate the time evolution of an $N$-body system is by using a composition of $2$-body problems to solve a more general $N$-body problem. While this approach may seem computationally expensive, our aim here is to present a theoretical formulation of the method. Possible optimizations, such as applying the Kepler-solver only to a few close pairs in the simulation, or to make use of Newton's third law during the force loop, are left for future implementations.

We first rewrite the potential energy term in eq.~\ref{eq:03} as follows:
\begin{eqnarray}
\nonumber H_{U} &=& \frac{1}{2}\sum_{i=1}^{N}\sum_{j\ne i}^{N} H_{U_{ij}}\\
\nonumber &=& \frac{1}{2}\sum_{i=1}^{N}\sum_{j\ne i}^{N}-\frac{m_{i} m_{j}}{r_{ij}}\\
\nonumber &=& \frac{1}{2}\sum_{i=1}^{N}\sum_{j\ne i}^{N}-\mu_{ij}\frac{(m_{i}+m_{j})}{r_{ij}}\\
\nonumber &=& \frac{1}{2}\sum_{i=1}^{N}\sum_{j\ne i}^{N}\mu_{ij}\left\{\left[\frac{1}{2}v^{2}_{ij} - \frac{(m_{i}+m_{j})}{r_{ij}}\right] - \frac{1}{2}v^{2}_{ij}\right\}\\
\nonumber &=& \frac{1}{2}\sum_{i=1}^{N}\sum_{j\ne i}^{N}\left(H_{K_{ij}} - H_{T_{ij}}\right)\\
&=& \frac{1}{2}\sum_{i=1}^{N}\sum_{j\ne i}^{N} H_{W_{ij}} \,\equiv\, H_{W}\,.\label{eq:04}
\end{eqnarray}
Here,
\begin{eqnarray}
H_{K_{ij}} \equiv \mu_{ij}\left[\frac{1}{2}v^{2}_{ij} - \frac{(m_{i}+m_{j})}{r_{ij}}\right]\,\label{eq:05}
\end{eqnarray}
is the $2$-body Keplerian Hamiltonian and
\begin{eqnarray}
H_{T_{ij}} \equiv \frac{1}{2}\mu_{ij}v^{2}_{ij},\label{eq:06}
\end{eqnarray}
where $\mu_{ij}=m_{i}m_{j}/(m_{i}+m_{j})$ is the reduced mass of the $i-j$ pair. The original $N$-body Hamiltonian in eq.~\ref{eq:01} can now be rewritten as follows:
\begin{eqnarray}
H = H_{T} + H_{W} = \sum_{i=1}^{N} H_{T_{i}} + \frac{1}{2}\sum_{i=1}^{N}\sum_{j\ne i}^{N} H_{W_{ij}}\,,\label{eq:07}
\end{eqnarray}
with
\begin{eqnarray}
H_{W_{ij}} = H_{K_{ij}} - H_{T_{ij}} \equiv H_{U_{ij}}\,.\label{eq:08}
\end{eqnarray}
We note that eq.~\ref{eq:07}, as is clear from the equivalence in eq.~\ref{eq:08}, can always be reduced by simplification into eq.~\ref{eq:01}, which implies that, in principle, our Keplerian-based Hamiltonian splitting does not change the dynamics of the system.

\subsection{Equations of motion}
\label{sec2.2:EoM}

According to the general theory of geometric integrators \citep{Hairer02} we can construct a time-symmetric second-order method by composing a (possible non-time-symmetric) first order method, $\phi(\tau)$, with its self-adjoint, $\phi^{\dagger}(\tau)$. Moreover, the composition $\Psi(\tau) = \phi(\frac{\tau}{2})\circ \phi^{\dagger}(\frac{\tau}{2})$ is symplectic if both $\phi(\tau)$ and $\phi^{\dagger}(\tau)$ are symplectic methods.

In our Keplerian-based Hamiltonian splitting, time evolution operators can be constructed by taking into account that
\begin{eqnarray}
e^{\tau \widehat{H}_{T}} = \prod_{i=1}^{N} e^{\tau \widehat{H}_{T_{i}}}\,,\label{eq:09}
\end{eqnarray}
\begin{eqnarray}
e^{\tau \widehat{H}_{U}} = \prod_{i=1}^{N} e^{\tau \frac{1}{2}\widehat{H}_{U_{i}}} = \prod_{i=1}^{N} e^{\tau \frac{1}{2}\sum_{j\ne i}^{N} \widehat{H}_{U_{ij}}} \,.\label{eq:10}
\end{eqnarray}
and, by eq.~\ref{eq:04},
\begin{eqnarray}
e^{\tau \widehat{H}_{W}} = \prod_{i=1}^{N} e^{\tau \frac{1}{2}\widehat{H}_{W_{i}}} = \prod_{i=1}^{N} e^{\tau \frac{1}{2}\sum_{j\ne i}^{N} \widehat{H}_{W_{ij}}} \,,\label{eq:11}
\end{eqnarray}
where the last term on the right hand side in eqs.~\ref{eq:10} and \ref{eq:11} is a simple substitution of the definition of operators $\widehat{H}_{U_{i}}$ and $\widehat{H}_{W_{i}}$, i.e., $\widehat{H}_{U_{i}} = \sum_{j\ne i}^{N} \widehat{H}_{U_{ij}}$ and similar for $\widehat{H}_{W_{i}}$, and the presence of the factor $1/2$ follows from the fact that we have to take into account each $i-j$ pair only once. In eqs.~\ref{eq:09}, \ref{eq:10} and \ref{eq:11} the individual operators $e^{\tau \widehat{H}_{T_{i}}}$, $e^{\tau \widehat{H}_{U_{i}}}$ and $e^{\tau \widehat{H}_{W_{i}}}$ act on the $6N$ dimensional array $(\mathbf{r}_{i}, \mathbf{v}_{i})$. Here the ``one-subscript'' operators individually commutate since they can only act on the corresponding coordinates with subscript $i$. Therefore, the order in which the product of operators is executed in each of eqs.~\ref{eq:09}, \ref{eq:10} and \ref{eq:11} is unimportant. In order to proceed with the derivation we present these operators in a more explicit form as follows:
\begin{eqnarray}
e^{\tau \widehat{H}_{T_{i}}}:\quad
\begin{pmatrix}
\mathbf{r}_{i}\\
\mathbf{v}_{i}
\end{pmatrix}
\leftarrow
\begin{pmatrix}
\mathbf{r}_{i}\\
\mathbf{v}_{i}
\end{pmatrix}
+ \tau
\begin{pmatrix}
\mathbf{v}_{i}\\
\mathbf{0}
\end{pmatrix}
\,,\label{eq:12}
\end{eqnarray}
\begin{eqnarray}
e^{\tau \widehat{H}_{U_{i}}}:\quad
\begin{pmatrix}
\mathbf{r}_{i}\\
\mathbf{v}_{i}
\end{pmatrix}
\leftarrow
\begin{pmatrix}
\mathbf{r}_{i}\\
\mathbf{v}_{i}
\end{pmatrix}
+ \tau
\begin{pmatrix}
\mathbf{0}\\
\mathbf{a}_{i}
\end{pmatrix}
\,,\label{eq:13}
\end{eqnarray}
\begin{eqnarray}
e^{\tau \widehat{H}_{W_{i}}}:\quad
\begin{pmatrix}
\mathbf{r}_{i}\\
\mathbf{v}_{i}
\end{pmatrix}
\leftarrow
\begin{pmatrix}
\mathbf{r}_{i}\\
\mathbf{v}_{i}
\end{pmatrix}
+
\begin{pmatrix}
\mathbf{\delta r}_{i}\\
\mathbf{\delta v}_{i}
\end{pmatrix}
\,,\label{eq:14}
\end{eqnarray}
where $\mathbf{a}_{i}$ is the acceleration and $(\delta\mathbf{r}_{i}, \delta\mathbf{v}_{i})$ are the increments in absolute coordinates and will be specified later on in eq.~\ref{eq:23}.

In a similar way, individual ``two-subscript'' operators are explicitly written as follows:
\begin{eqnarray}
e^{\tau \widehat{H}_{U_{ij}}}:\quad
\begin{pmatrix}
\mathbf{r}_{ij}\\
\mathbf{v}_{ij}
\end{pmatrix}
\leftarrow
\begin{pmatrix}
\mathbf{r}_{ij}\\
\mathbf{v}_{ij}
\end{pmatrix}
+ \tau
\begin{pmatrix}
\mathbf{0}\\
\mathbf{a}_{ij}
\end{pmatrix}
\,,\label{eq:15}
\end{eqnarray}
\begin{eqnarray}
e^{\tau \widehat{H}_{W_{ij}}}:\quad
\begin{pmatrix}
\mathbf{r}_{ij}\\
\mathbf{v}_{ij}
\end{pmatrix}
\leftarrow
\begin{pmatrix}
\mathbf{r}_{ij}\\
\mathbf{v}_{ij}
\end{pmatrix}
+
\begin{pmatrix}
\mathbf{\delta r}_{ij}\\
\mathbf{\delta v}_{ij}
\end{pmatrix}
\,,\label{eq:16}
\end{eqnarray}
\begin{eqnarray}
e^{\tau \left(-\widehat{H}_{T_{ij}}\right)}:\quad
\begin{pmatrix}
\mathbf{r}_{ij}\\
\mathbf{v}_{ij}
\end{pmatrix}
\leftarrow
\begin{pmatrix}
\mathbf{r}_{ij}\\
\mathbf{v}_{ij}
\end{pmatrix}
- \tau
\begin{pmatrix}
\mathbf{v}_{ij}\\
\mathbf{0}
\end{pmatrix}
\,,\label{eq:17}
\end{eqnarray}
\begin{eqnarray}
e^{\tau \widehat{H}_{K_{ij}}}:\quad
\mathbf{r}_{ij}, \mathbf{v}_{ij} \leftarrow \mathtt{kepler\_solver}(\tau, m_{ij}, \mathbf{r}_{ij}, \mathbf{v}_{ij})\,,\label{eq:18}
\end{eqnarray}
where $m_{ij} = m_{i} + m_{j}$, $\mathbf{a}_{ij}=-m_{ij}\mathbf{r}_{ij}/r_{ij}^{3}$ is the relative $2$-body acceleration. The increments in relative coordinates, $(\delta\mathbf{r}_{ij}, \delta\mathbf{v}_{ij})$, are obtained independently for each $i-j$ pair from the application of one of the first-order maps:
\begin{subequations}
\begin{eqnarray}
e^{\tau \left(\widehat{H}_{K_{ij}} - \widehat{H}_{T_{ij}}\right)} \approx e^{\tau \left(-\widehat{H}_{T_{ij}}\right)} e^{\tau \widehat{H}_{K_{ij}}}\,,\label{eq:19a}\\
e^{\tau \left(\widehat{H}_{K_{ij}} - \widehat{H}_{T_{ij}}\right)} \approx e^{\tau \widehat{H}_{K_{ij}}} e^{\tau \left(-\widehat{H}_{T_{ij}}\right)}\,.\label{eq:19b}
\end{eqnarray}
\end{subequations}

Eqs.~\ref{eq:12} to \ref{eq:17} are first-order approximations to the respective operators in these equations. It will be clear below that this low-order approximation is enough for our purposes since, ultimately, the order of the full time evolution operator in eq.~\ref{eq:25} will be determined by the composition of those operators. In this sense, if a high-order approximation of the method presented here is needed, we argue that this should be obtained not by extending eqs.~\ref{eq:12} to \ref{eq:17} to higher order, but rather, by making a high-order composition of these operators in a similar way as in symplectic integrators \citep{Yoshida90,Kinoshita_etal91}, where a second-order map is constructed as a composition of first-order operators, and so on.

We notice here that, contrary to the ``one-subscript'' operators, the ``two-subscript'' operators act on the $6N(N-1)/2$ dimensional array $(\mathbf{r}_{ij}, \mathbf{v}_{ij})$. Therefore, it remains to be shown how to relate ``one-subscript'' and ``two-subscript'' operators in a consistent way. From eq.~\ref{eq:10} and the definition of $H_{U_{ij}}$, it is easy to see that the equivalence,
\begin{eqnarray}
\prod_{j\ne i}^{N} e^{\tau \widehat{H}_{U_{ij}}} \equiv e^{\tau \sum_{j\ne i}^{N} \widehat{H}_{U_{ij}}} = e^{\tau \widehat{H}_{U_{i}}}\,,\label{eq:20}
\end{eqnarray}
is valid for every $N$ because the operators $e^{\tau \widehat{H}_{U_{ij}}}$ commutate. On the other hand, from eq.~\ref{eq:11}, an equivalence similar to eq.~\ref{eq:20} relating $\widehat{H}_{W}$-type operators is only possible for $N = 2$. For $N > 2$ the operators $e^{\tau \widehat{H}_{W_{ij}}}$ do not commutate. However, we can write a similar equation approximately as
\begin{eqnarray}
\prod_{j\ne i}^{N} e^{\tau \widehat{H}_{W_{ij}} + \mathcal{O}(\tau^2)} \approx e^{\tau \sum_{j\ne i}^{N} \widehat{H}_{W_{ij}}} = e^{\tau \widehat{H}_{W_{i}}}\,,\label{eq:21}
\end{eqnarray}
where the error $\mathcal{O}(\tau^2)$ is not guaranteed to be Hamiltonian due to the fact that we treat each $i-j$ pair independently. As a consequence \textit{the symplecticity of the present method is lost}.

Apart from the loss of symplecticity, as mentioned above, a time-symmetric second-order method for our Keplerian-based Hamiltonian splitting can still be constructed by using a composition of self-adjoint first-order methods (see \cite{Hairer02}).

In order to construct $\phi(\tau)$ and $\phi^{\dagger}(\tau)$ we first need to specify the increments $\delta\mathbf{r}_{i}$ and $\delta\mathbf{v}_{i}$ in eq.~\ref{eq:14}. Since in the present method we take advantage of a \texttt{kepler\_solver} to evolve each pair of particles independently, the relative increments $(\delta\mathbf{r}_{ij}, \delta\mathbf{v}_{ij})$ can be easily calculated for each interaction after application of one of the maps in eqs.~\ref{eq:19a} or \ref{eq:19b}. Here, what we seek is an approximate relation between the increments in relative coordinates $(\delta\mathbf{r}_{ij}, \delta\mathbf{v}_{ij})$ and those in absolute coordinates $(\delta\mathbf{r}_{i}, \delta\mathbf{v}_{i})$, in order to construct the full integrator. By noting that increments associated with operators $\widehat{H}_{U_{ij}}$ and $\widehat{H}_{U_{i}}$ are related by
\begin{eqnarray}
\tau
\begin{pmatrix}
\mathbf{0}\\
\mathbf{a}_{i}
\end{pmatrix}
= \frac{1}{m_{i}} \sum_{j\ne i}^{N} \mu_{ij} \tau
\begin{pmatrix}
\mathbf{0}\\
\mathbf{a}_{ij}
\end{pmatrix}
\,,\label{eq:22}
\end{eqnarray}
a way to specify $(\delta\mathbf{r}_{i}, \delta\mathbf{v}_{i})$ consists of exploring the equivalence between $H_{U}$ and $H_{W}$, as first presented in eq.~\ref{eq:04}. In addition, if we take into account the discussion above regarding to eqs.~\ref{eq:10}, \ref{eq:11}, \ref{eq:20} and \ref{eq:21}, a relation between relative and absolute increments can be defined in analogy to eq.~\ref{eq:22} as follows:
\begin{eqnarray}
\begin{pmatrix}
\delta \mathbf{r}_{i}\\
\delta \mathbf{v}_{i}
\end{pmatrix}
= \frac{1}{m_{i}} \sum_{j\ne i}^{N} \mu_{ij}
\begin{pmatrix}
\delta \mathbf{r}_{ij}\\
\delta \mathbf{v}_{ij}
\end{pmatrix}
+ \mathcal{O}(\tau^2)
\,,\label{eq:23}
\end{eqnarray}
which constitutes a first-order approximation as explained above (see eq.~\ref{eq:21}). While we were not able to provide a more formal derivation to eq.~\ref{eq:23}, we will show below (see explanation about eq.~\ref{eq:28}) that when we calculate the relative increments from an ordinary Leapfrog map rather than the \texttt{kepler\_solver} in eq.~\ref{eq:18}, then eq.~\ref{eq:23} reduces to eq.~\ref{eq:22}.

We can now define a time-symmetric second-order map for our Keplerian-based Hamiltonian splitting as follows:
\begin{eqnarray}
\nonumber \Psi(\tau) &\equiv& \phi(\frac{\tau}{2})\circ \phi^{\dagger}(\frac{\tau}{2})\,,\\
&\equiv& e^{\frac{\tau}{2} \widehat{H}_{T}} e^{\frac{\tau}{2} \widehat{H}_{W}} \circ e^{\frac{\tau}{2} \widehat{H}_{W}} e^{\frac{\tau}{2} \widehat{H}_{T}}\,,\label{eq:24}
\end{eqnarray}
where the increments $(\delta\mathbf{r}_{ij}, \delta\mathbf{v}_{ij})$ which appear in the $e^{\frac{\tau}{2} \widehat{H}_{W}}$ operator on the left side of $\circ$ are independently obtained after application of eq.~\ref{eq:19a} for each $i-j$ pair, while those which appear on the right side of $\circ$ are independently obtained after application of the (self-adjoint) method in eq.~\ref{eq:19b} for each $i-j$ pair. Eq.~\ref{eq:24} can be further simplified by merging operators on both sides of $\circ$, giving,
\begin{eqnarray}
\Psi(\tau) \equiv e^{\frac{\tau}{2} \widehat{H}_{T}} e^{\tau \widehat{H}_{W}} e^{\frac{\tau}{2} \widehat{H}_{T}}\,,\label{eq:25}
\end{eqnarray}
in which case the increments $(\delta\mathbf{r}_{ij}, \delta\mathbf{v}_{ij})$ appearing in the $e^{\tau \widehat{H}_{W}}$ operator should be independently obtained after application of a time-symmetric second-order map for each $i-j$ pair,
\begin{eqnarray}
e^{\tau \left(\widehat{H}_{K_{ij}} - \widehat{H}_{T_{ij}}\right)} \approx e^{\frac{\tau}{2} \left(-\widehat{H}_{T_{ij}}\right)} e^{\tau \widehat{H}_{K_{ij}}} e^{\frac{\tau}{2} \left(-\widehat{H}_{T_{ij}}\right)}\,.\label{eq:26}
\end{eqnarray}

The equations of motion that result from the full map in eq.~\ref{eq:25} can be written in the following discrete form:
\begin{subequations}
\label{eq:27}
\begin{eqnarray}
\mathbf{r}_{i}^{1/2} &=& \mathbf{r}_{i}^{0} + \frac{\tau}{2}\mathbf{v}_{i}^{0}\,,\label{eq:27a}\\
\tilde{\mathbf{r}}_{i} &=& \mathbf{r}_{i}^{1/2} + \frac{1}{m_{i}}\sum_{j\ne i}^{N}\mu_{ij} \delta\mathbf{r}_{ij}\,,\label{eq:27b}\\
\mathbf{v}_{i}^{1} &=& \mathbf{v}_{i}^{0} + \frac{1}{m_{i}}\sum_{j\ne i}^{N}\mu_{ij} \delta\mathbf{v}_{ij}\,,\label{eq:27c}\\
\mathbf{r}_{i}^{1} &=& \tilde{\mathbf{r}}_{i} + \frac{\tau}{2}\mathbf{v}_{i}^{1}\,,\label{eq:27d}
\end{eqnarray}
\end{subequations}
where $\mathbf{r}_{i}^{1}=\mathbf{r}_{i}(t+\tau)$, $\mathbf{r}_{i}^{0}=\mathbf{r}_{i}(t)$ and similar for $\mathbf{v}_{i}$, and the increments $(\delta\mathbf{r}_{ij}, \delta\mathbf{v}_{ij})$ are calculated independently as explained above.

As it can be seen, eqs.~\ref{eq:27} are remarkably similar to the Leapfrog method. It remains to be shown that these equations effectively reduce to the Leapfrog equations when we substitute the $2$-body \texttt{kepler\_solver} to a simple DKD-type integrator. In this case, the map in eq.~\ref{eq:26} becomes:
\begin{subequations}
\label{eq:28}
\begin{eqnarray}
\mathbf{r}_{ij} &\leftarrow& \mathbf{r}_{ij} - \frac{\tau}{2} \mathbf{v}_{ij}\,,\label{eq:28a}\\
\mathbf{r}_{ij} &\leftarrow& \mathbf{r}_{ij} + \frac{\tau}{2} \mathbf{v}_{ij}\,,\label{eq:28b}\\
\mathbf{v}_{ij} &\leftarrow& \mathbf{v}_{ij} + \tau \, \mathbf{a}_{ij}\,,\label{eq:28c}\\
\mathbf{r}_{ij} &\leftarrow& \mathbf{r}_{ij} + \frac{\tau}{2} \mathbf{v}_{ij}\,,\label{eq:28d}\\
\mathbf{r}_{ij} &\leftarrow& \mathbf{r}_{ij} - \frac{\tau}{2} \mathbf{v}_{ij}\,,\label{eq:28e}
\end{eqnarray}
\end{subequations}
which results in $\delta\mathbf{v}_{ij} = \tau \mathbf{a}_{ij}$ and $\delta\mathbf{r}_{ij} = \mathbf{0}$ and, in view of eqs.~\ref{eq:22} and \ref{eq:23}, completes the demonstration. It should be noted that in this particular case, the error in eq.~\ref{eq:23} disappears because $\delta\mathbf{r}_{ij} = \mathbf{0}$ and eq.~\ref{eq:21} reduces to eq.~\ref{eq:20}, restoring the symplecticity of the method.
Note also that this is true only if we use a DKD-type integrator as a $2$-body solver. For a KDK-type $2$-body solver the symplecticity of the method is not restored because the order in which $(\mathbf{r}_{ij}, \mathbf{v}_{ij})$ is evolved in eqs.~\ref{eq:28} changes and $\delta\mathbf{r}_{ij} \neq \mathbf{0}$. In other words, using a simple DKD-type integrator as a $2$-body solver in the scheme above results in a very expensive implementation of a traditional Leapfrog method.

On the other hand, with the \texttt{kepler\_solver} function as a $2$-body solver, a non-Hamiltonian error is made due to the non-commutativity of the $e^{\tau \widehat{H}_{W_{ij}}}$ operators and the fact that each $i-j$ pair is treated independently, leading to the loss of symplecticity of the resulting method. Because our Keplerian-based integrator is constructed as a composition of self-adjoint first-order maps, it still preserves time-reversibility and second-order convergence (error $\mathcal{O}(\tau^3)$).

The advantage of using the \texttt{kepler\_solver} instead, comes from the fact that it is guaranteed that all pairwise interactions are always integrated precisely, which, in practical $N$-body simulations, is a much stronger requirement than the symplecticity of the Hamiltonian flow.

\subsection{Implementation}
\label{sec2.3:implementation}

The method described in the previous section has been implemented in a new code called \texttt{Sakura}, which is available in Astrophysical MUlti-purpose Software Environment (\texttt{AMUSE}\footnote{www.amusecode.org}, \cite{PortegiesZwart_etal13}). In order to clarify the implementation, Listing~\ref{listing:01} shows a \texttt{Python}\footnote{The actual implementation has been done in \texttt{C/C++} for efficiency purposes.} code for the main loop calculation which evolves the particle's coordinates according to the map in eq.~\ref{eq:25} or, equivalently, eqs.~\ref{eq:27}. The \texttt{kepler\_solver} function at line $47$ implements a universal variable Kepler-solver closely following \cite{Conway86}. Note that the memory and CPU requirements of this code scales as $O(N)$ and $O(N^{2})$, respectively.
\begin{figure}
\begin{lstlisting}[caption={\texttt{Python} code for the main loop in \texttt{Sakura} integrator},label=listing:01,frame=single, basicstyle=\ttfamily\small, numbers=left, numberstyle=\tiny, stepnumber=1, numbersep=5pt]
"""The functions below implement the main
steps of Sakura integrator.

The required parameters are the following:

:param tau: the time-step size.
:param n: the number of particles.
:param m: array with particles' masses.
:param r: 3D array with particles' positions.
:param v: 3D array with particles' velocities.
"""

def do_step(tau, n, m, r, v):
  r, v = evolve_HT(tau/2, n, m, r, v)
  r, v = evolve_HW(tau, n, m, r, v)
  r, v = evolve_HT(tau/2, n, m, r, v)
  return r, v

def evolve_HT(tau, n, m, r, v):
  for i in range(n):
    for k in range(3):
      r[i][k] += v[i][k] * tau
  return r, v

def evolve_HW(tau, n, m, r, v):
  # Allocate/initialize 3D arrays to store
  # increments in position/velocity due to
  # 2-body interactions.
  dmr = numpy.zeros((n, 3))
  dmv = numpy.zeros((n, 3))
  
  # For each i-j pair, this corresponds to
  # the eq. 26 in the main text.
  for i in range(n):
    for j in range(n):
      if i != j:
        mij = m[i] + m[j]
        mu = m[i] * m[j] / mij
        for k in range(3):
          rr0[k] = r[i][k] - r[j][k]
          vv0[k] = v[i][k] - v[j][k]
        ###
        for k in range(3):
          r0[k] = rr0[k] - vv0[k] * tau / 2
          v0[k] = vv0[k]
        #
        r1, v1 = kepler_solver(tau, mij, r0, v0)
        #
        for k in range(3):
          rr1[k] = r1[k] - v1[k] * tau / 2
          vv1[k] = v1[k]
        ###
        for k in range(3):
          dmr[i][k] += mu * (rr1[k] - rr0[k])
          dmv[i][k] += mu * (vv1[k] - vv0[k])
          
  # This corresponds to eqs. 27b and 27c
  # in the main text.
  for i in range(n):
    for k in range(3):
      r[i][k] += dmr[i][k] / m[i]
      v[i][k] += dmv[i][k] / m[i]
  return r, v
\end{lstlisting}
\end{figure}

\section{Tests}
\label{sec3:tests}

In order to verify that \texttt{Sakura} performs well on collisional $N$-body systems, we present some tests for $N$ ranging from a few to a thousand. We compare the results of \texttt{Sakura} to those obtained using a modified version of the Leapfrog integrator and a standard $4$-th order Hermite integrator, available in the \texttt{AMUSE} framework. The modification in the Leapfrog integrator consists of the introduction of a routine to allow the use of adaptive time-steps. In this case the time-symmetry of the Leapfrog method is still preserved because we adopted the recipe for time-symmetrization as suggested in \cite{Pelupessy_etal12}. A comparison of the computational costs and scalings with $N$ is also presented. We emphasize that the base time-step size in each of the tests of \texttt{Sakura} is kept constant during the simulation, whilst in Leapfrog and Hermite integrations a shared adaptive time-step scheme has been adopted. The time-step criterion used within Leapfrog integrations is the time-symmetrized version of $\tau \sim \min((r_{ij}/a_{ij})^{1/2})$, whilst in Hermite code the standard Aarseth-criterion is used. For other details about these codes we refer the reader to the \texttt{AMUSE} documentation\footnotemark[3]. The value of the constant time-step size in \texttt{Sakura} is chosen in such a way that the same number of integration steps is taken as in the case of the Hermite integrations. Similarly, the time-step parameter in Leapfrog integrations is chosen to give approximately the same number of steps as in Hermite integrations. Note that, by construction, \texttt{Sakura} does not admit any softening parameter. Therefore, we also use zero softening in the other methods.

\subsection{Small-$N$ systems}
\label{sec3.1:smalln}

We start by presenting some numerical tests for well known simple small-$N$ systems including the figure-eight system ($N=3$; \cite{ChencinerMontgomery2000}), the Pythagorean system ($N=3$; \cite{SzebehelyPeters1967}) and the Sun with planets\footnote{We include Pluto in our simulations of the solar system since we use the initial conditions as given in \cite{ItoTanikawa02}.} ($N=10$; \cite{ItoTanikawa02}). We do not show results for a single binary system ($N=2$), since in this case \texttt{Sakura} reduces to an ordinary Kepler-solver which gives a solution for the binary orbit accurate to machine precision. The simulation time spans 100 $N$-body units \citep{HeggieMathieu1986} in the case of the first two systems and $10^{3}\,\mathrm{yr}$ in the case of the solar system.

\begin{figure*}
\centering
% figure8
\includegraphics[scale=0.8]{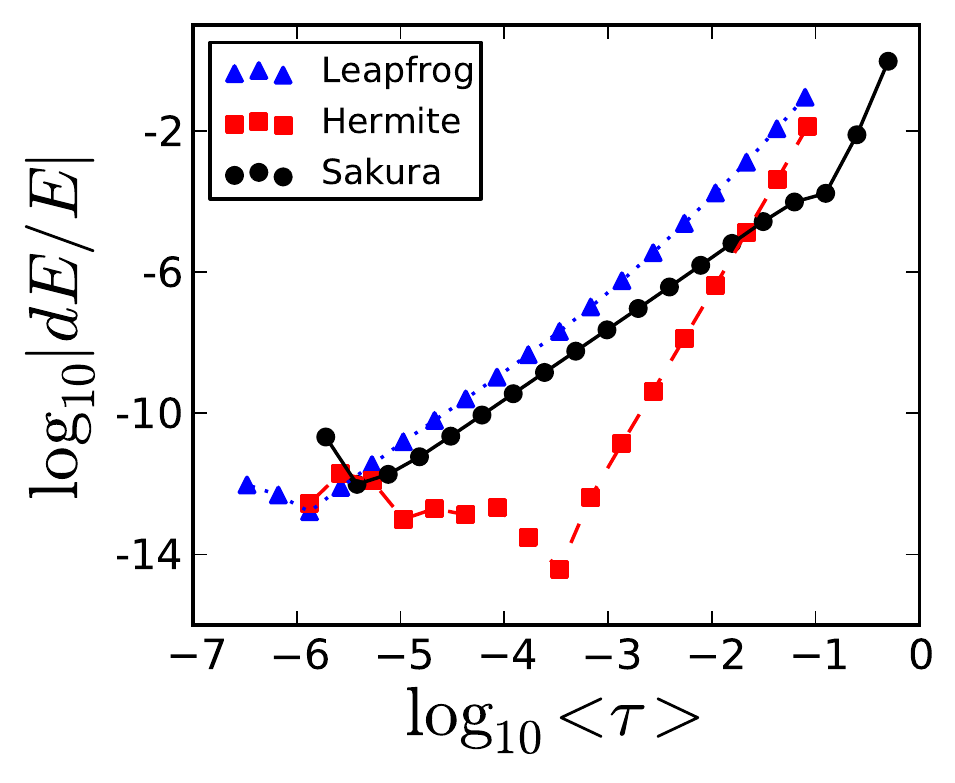}
\includegraphics[scale=0.8]{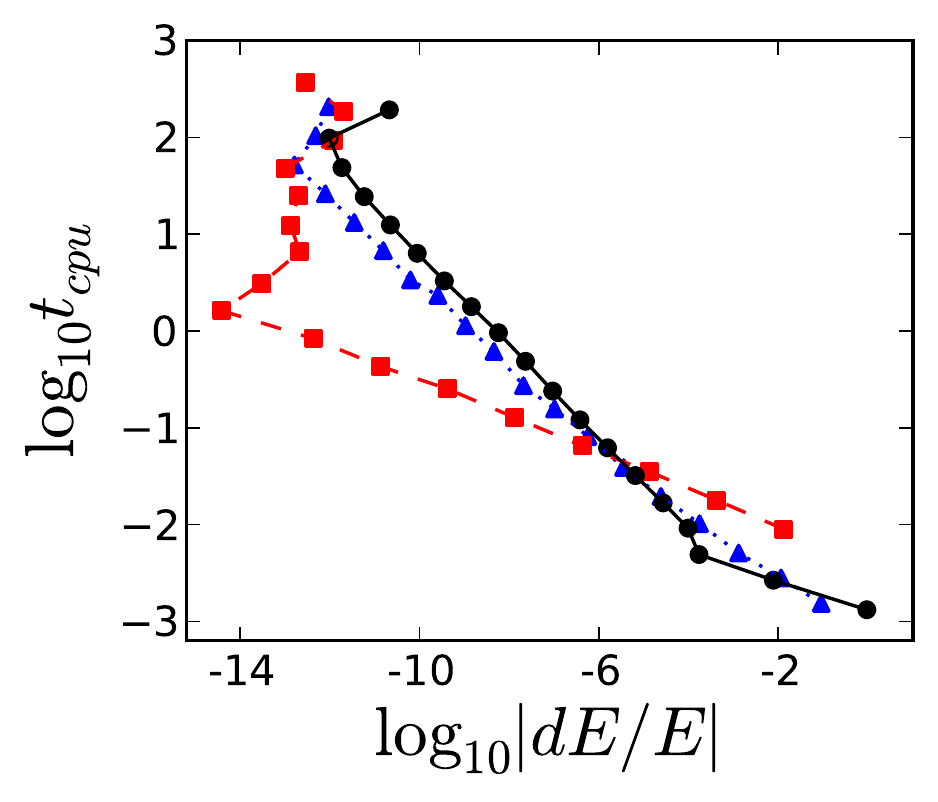}
% pythagorean
\includegraphics[scale=0.8]{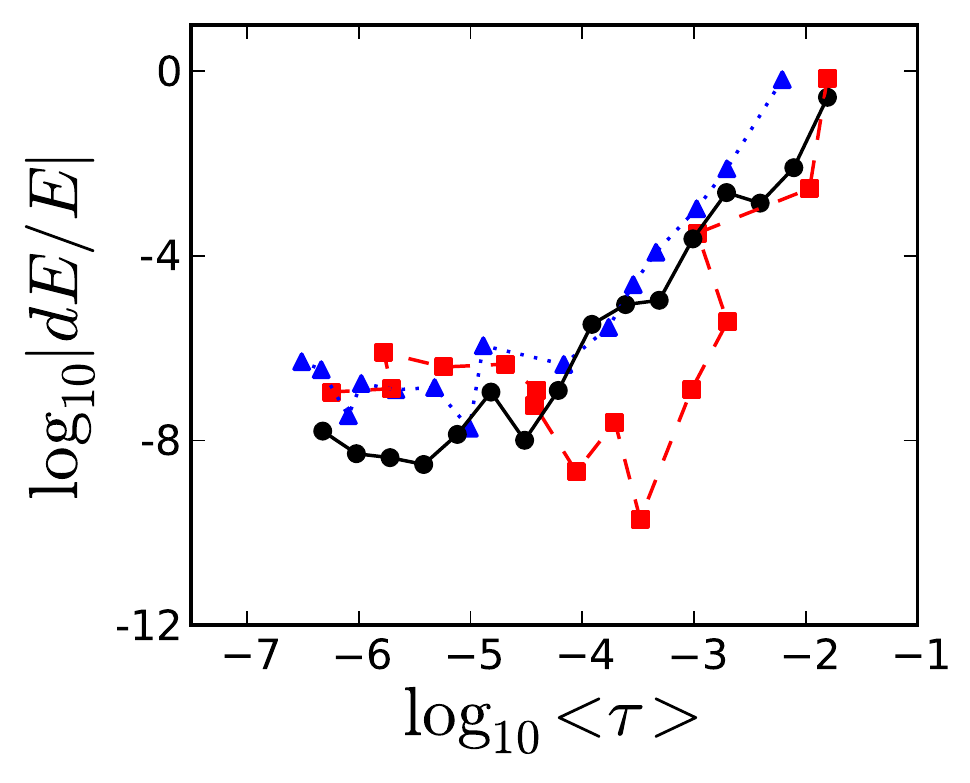}
\includegraphics[scale=0.8]{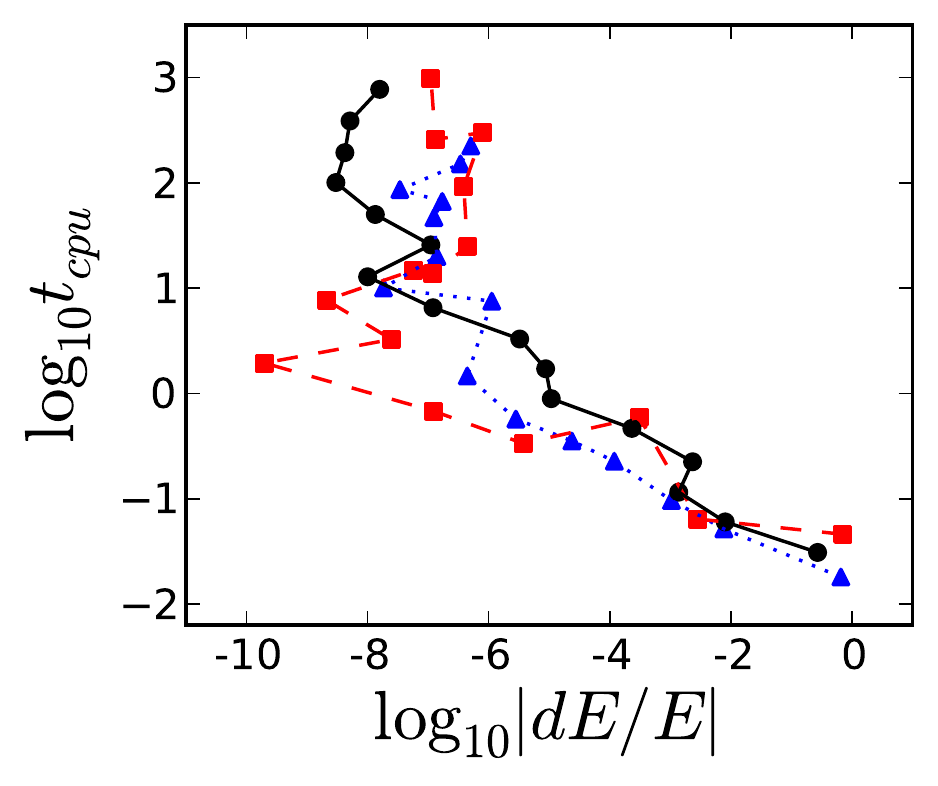}
% solar-system
\includegraphics[scale=0.8]{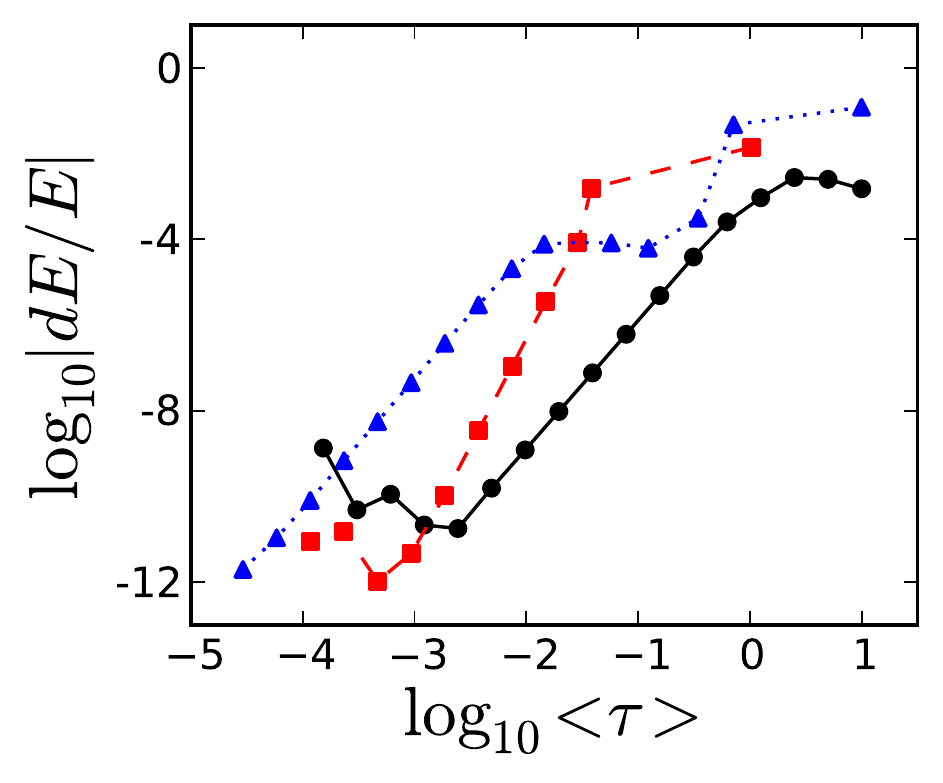}
\includegraphics[scale=0.8]{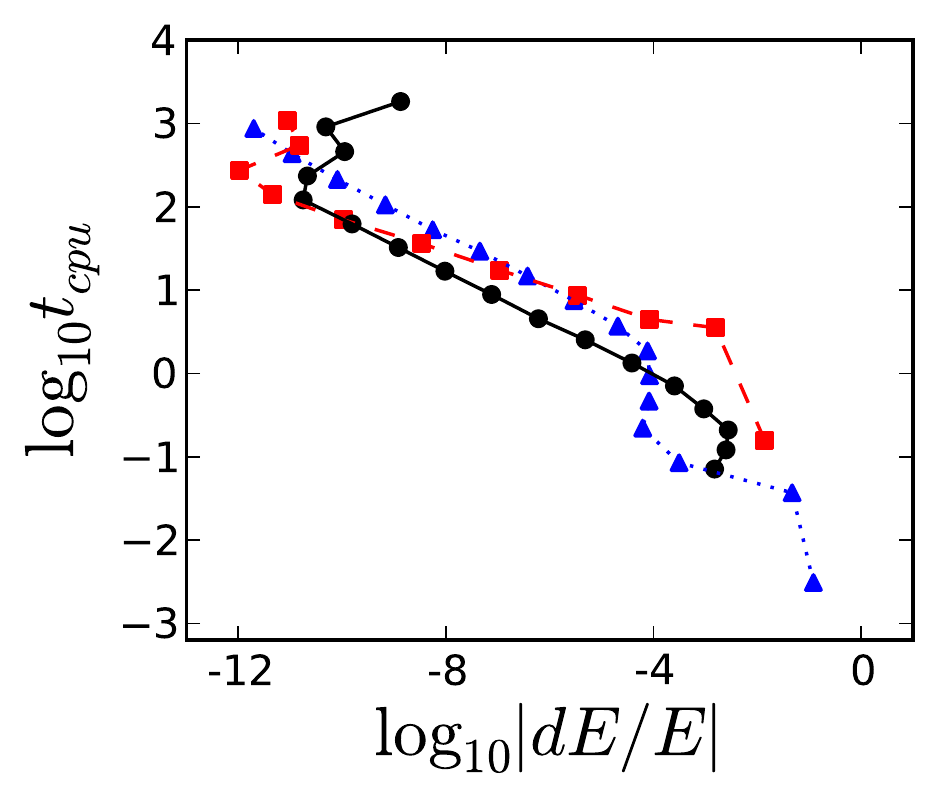}
\caption{Relative energy error as a function of the average time-step size (left panels) and CPU time (in seconds) vs relative energy error (right panels) for the Leapfrog integrator (blue triangles), $4$-th order Hermite (red squares) and \texttt{Sakura} (black bullets) for three different systems: figure-eight system (top panels), Pythagorean system (middle panels) and Sun with planets (bottom panels). $\langle \tau \rangle$ is given in $N$-body units and stands for the average value of the shared adaptive time-step size in Hermite integrations.}
\label{fig:01}
\end{figure*}

In Fig.~\ref{fig:01}, we present the relative energy error as a function of the average time-step size (left panels) and CPU time vs relative energy error (right panels) for the figure-eight system (top panels), Pythagorean system (middle panels) and Sun with planets (bottom panels), for the Leapfrog, $4$-th order Hermite and \texttt{Sakura}. We note that for the figure-eight system the $4$-th order Hermite usually performs better than Leapfrog and \texttt{Sakura} for a level of energy conservation $\lesssim 10^{-6}$. We attribute this to the fact that in this system the intrinsic time-step size of the particles does not change considerably during the orbital evolution and then, for smaller $\tau$, the $4$-th order convergence rate of the Hermite integrator outperforms Leapfrog and \texttt{Sakura}, which are of $2$-nd order. We notice that in this case, where all three particles democratically interact among themselves, \texttt{Sakura} is not expected to be the most suitable method of integration due to the non-commutativity of $2$-body interactions. Nevertheless, as we see in Fig.~\ref{fig:01} (top panels), its performance is comparable to that of the Leapfrog integrator. For the Pythagorean system, which contains several close encounters between particles during its orbital evolution, all three integration methods are somewhat comparable, despite \texttt{Sakura} using constant time-steps and the other two methods using adaptive time-steps. For the solar-system, in which the orbital evolution of the planets is almost Keplerian, \texttt{Sakura} delivers about $4$ orders of magnitude better energy conservation than Leapfrog, being also more precise than Hermite integration for time-steps $\gtrsim 10^{-3}$, while consuming the least amount of CPU time.

\begin{figure*}
\centering
\includegraphics[scale=0.8]{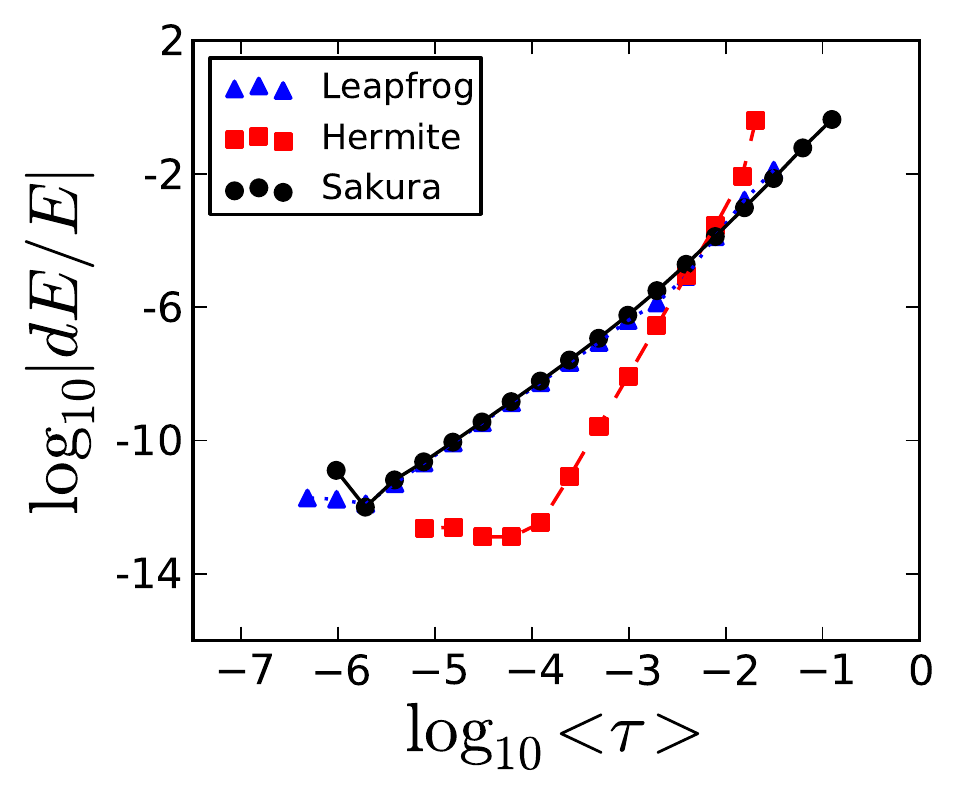}
\includegraphics[scale=0.8]{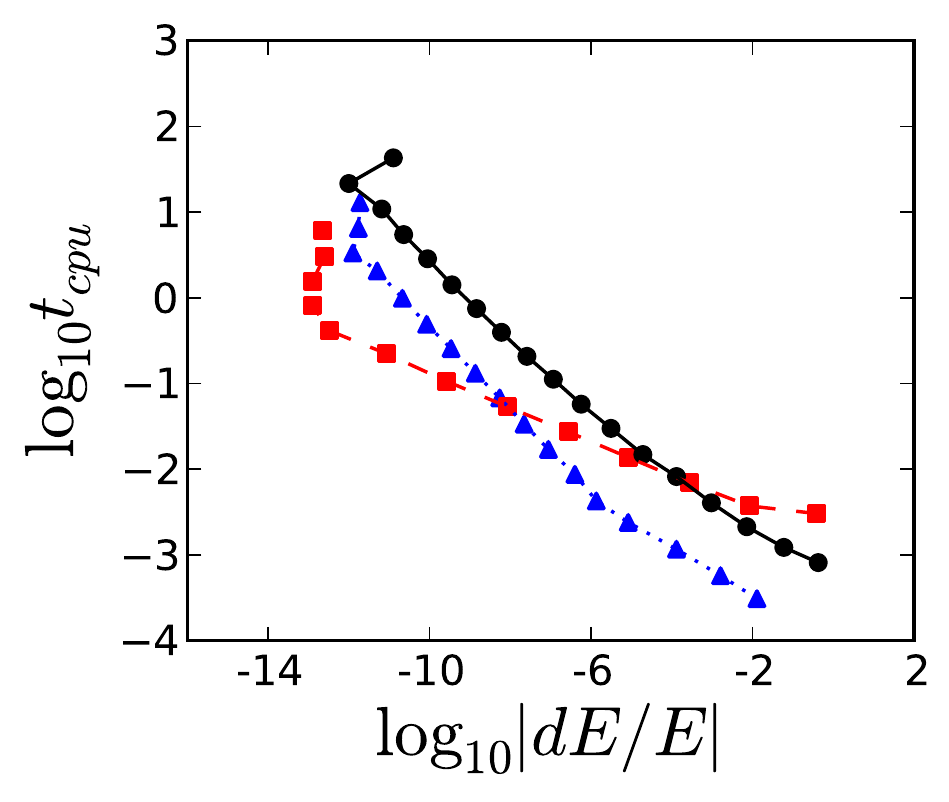}
\includegraphics[scale=0.8]{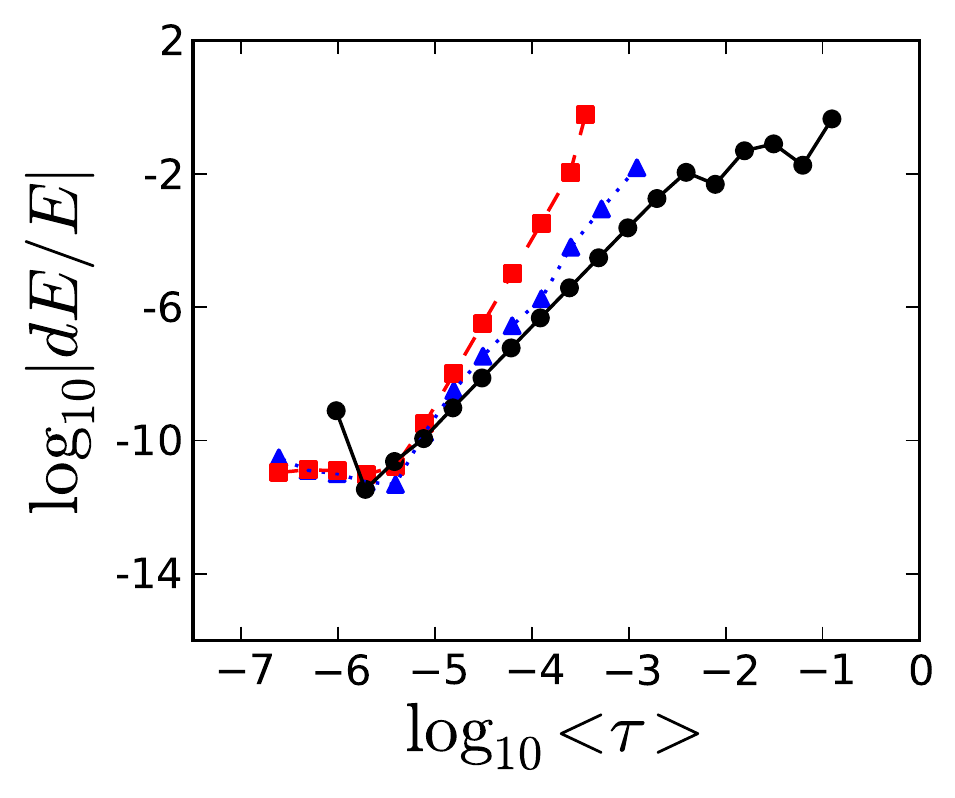}
\includegraphics[scale=0.8]{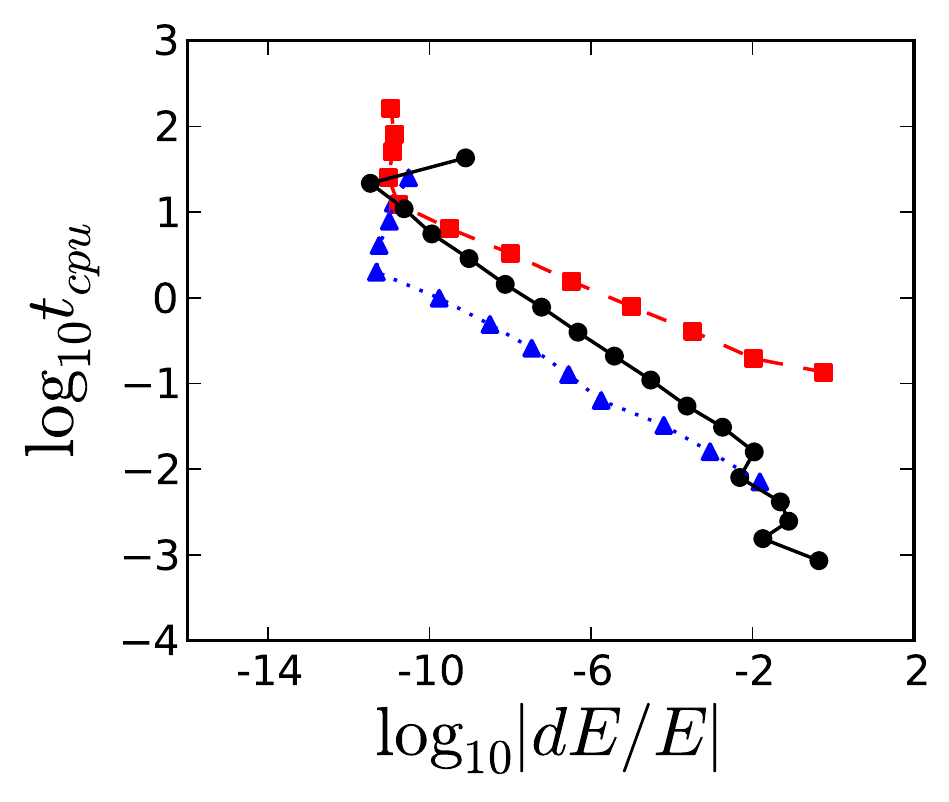}
\includegraphics[scale=0.8]{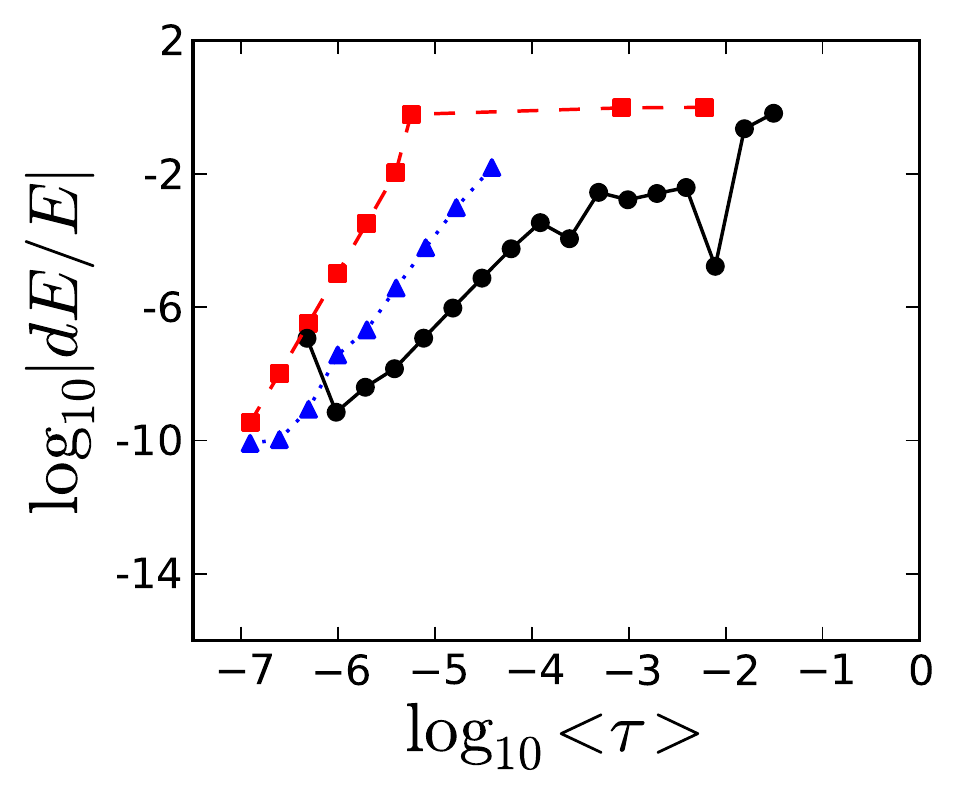}
\includegraphics[scale=0.8]{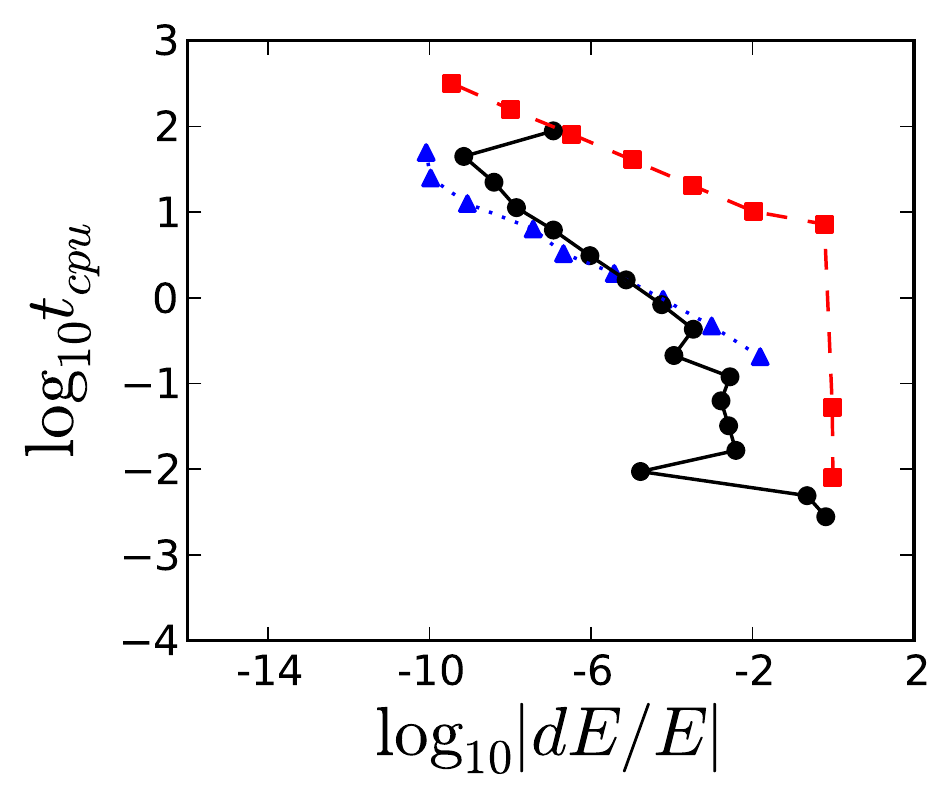}
\caption{The same as Fig.~\ref{fig:01} but for the hierarchical binary system for the following semi-major axis ratios: $a_{\mathrm{outer}}/a_{\mathrm{inner}}=10$ (top panels), $a_{\mathrm{outer}}/a_{\mathrm{inner}}=100$ (middle panels), $a_{outer}/a_{inner}=1000$ (bottom panels).}
\label{fig:02}
\end{figure*}

For those kind of systems, an integration step using \texttt{Sakura} is usually more expensive than an integration step using Hermite or Leapfrog by a factor $2-4$. Also, since all these codes scales as $O(N^{2})$, these figures are expected to remain unchanged when the number of particles increases. However, due to the fact that \texttt{Sakura} can handle compact binaries and/or resolve close encounters even with constant $\tau$, less time-steps are required for a given level of energy conservation implying that in these cases \texttt{Sakura} might outperform Hermite and Leapfrog integrations. In order to confirm this, we also include a test with a specially constructed initial condition which consists of a hierarchical binary system ($N=4$) with two tight binaries orbiting around each other in a circular orbit with semi-major axis $a_{\mathrm{outer}}=1$ ($N$-body units). The particles in each tight binary are themselves in a circular orbit with semi-major axis $a_{\mathrm{inner}}$. We have selected a semi-major axis ratio in the range $a_{\mathrm{outer}}/a_{\mathrm{inner}}=10-1000$, and performed a simulation for these systems for a time span of one $P_{\mathrm{outer}}$, i.e., the largest orbital period in the system (which is the same for all semi-major axis ratios). In Fig.~\ref{fig:02} we present the relative energy error as a function of the time-step size (left panels) and CPU time vs relative energy error (right panels) for the $4$-th order Hermite, Leapfrog and \texttt{Sakura}.

For the $a_{\mathrm{outer}}/a_{\mathrm{inner}}=10$ case (top panels in Fig.~\ref{fig:02}), \texttt{Sakura} delivers the same level of energy conservation as Leapfrog, although being more time consuming, whilst $4$-th order Hermite has better energy conservation due to its higher order convergence for time-step sizes $\lesssim 10^{-2}$. However, for tighter interacting binaries (middle and bottom panels in Fig.~\ref{fig:02}), \texttt{Sakura} shows increasingly better performance with the compactness of the interacting binaries. In particular, for a level of energy conservation of $10^{-6}$, typically adopted in collisional $N$-body simulations, \texttt{Sakura} is more than a order of magnitude faster than Hermite for the tightest binary configuration, $a_{\mathrm{outer}}/a_{\mathrm{inner}}=1000$, while having a similar speed as Leapfrog. Also for the tightest binary configuration, \texttt{Sakura} is the most precise integration method for a range in time-steps of $6$ orders of magnitude. On the other hand, for this latter system, the $4$-th order Hermite results only start converging to good energy conservation when using time-steps $\lesssim 10^{-5.5}$, which in some circumstances might be impractical in computational terms, when systems of this kind are present in a large-scale simulation.

\subsection{Large-$N$ systems}
\label{sec3.2:largen}

\begin{figure*}
\centering
\includegraphics[scale=0.8]{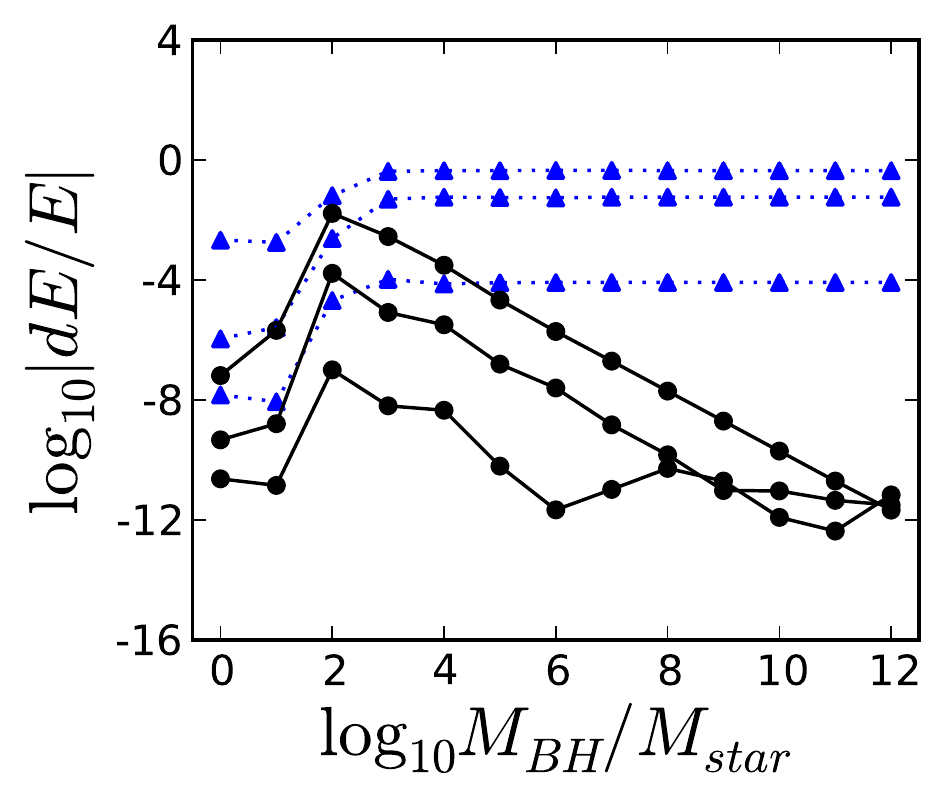}
\includegraphics[scale=0.8]{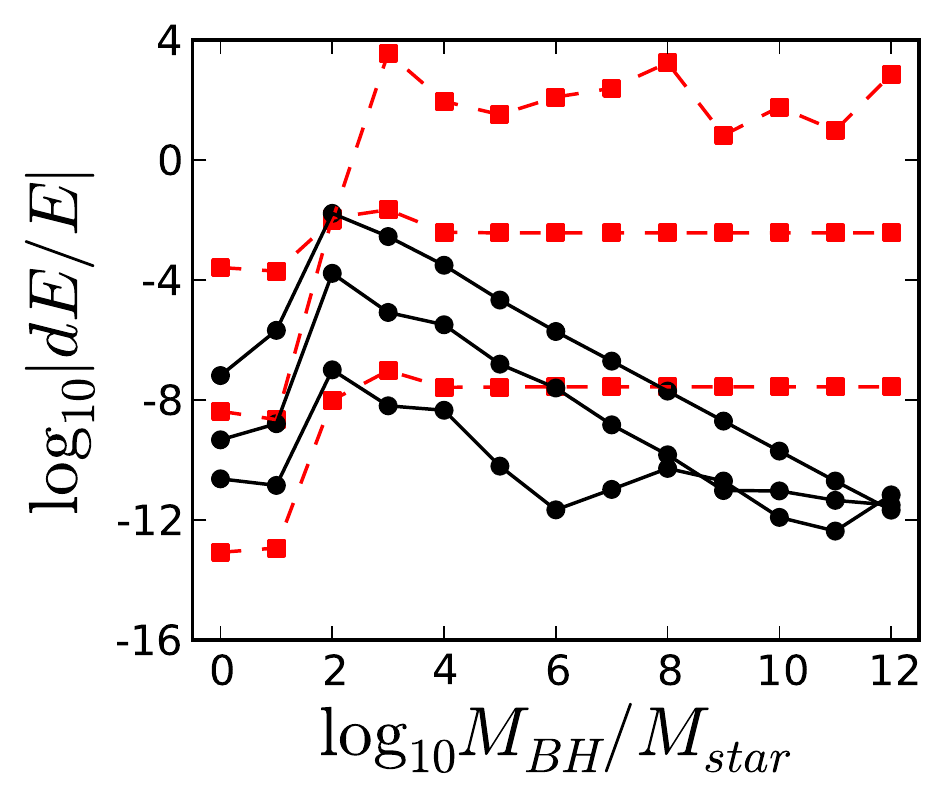}
\caption{For a Plummer sphere with a central black-hole, the panels show a comparison of the relative energy error as a function of the black-hole to stellar mass ratio for time-step sizes $\langle \tau \rangle = 10^{-3}, 10^{-4}, 10^{-5}$ (top, middle and bottom lines). The left panel present the results for Leapfrog (blue triangles) and \texttt{Sakura} (black bullets) and the right panel present the results for $4$-th order Hermite (red squares) and \texttt{Sakura} (black bullets).}
\label{fig:03}
\end{figure*}

\begin{figure*}
\centering
\includegraphics[scale=0.8]{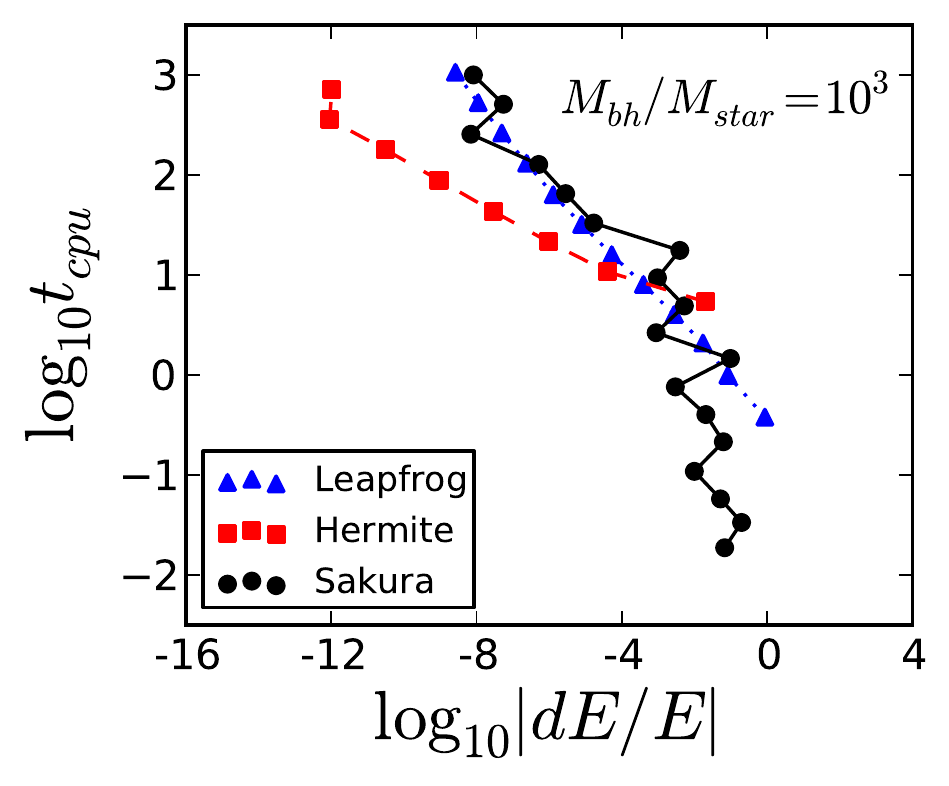}
\includegraphics[scale=0.8]{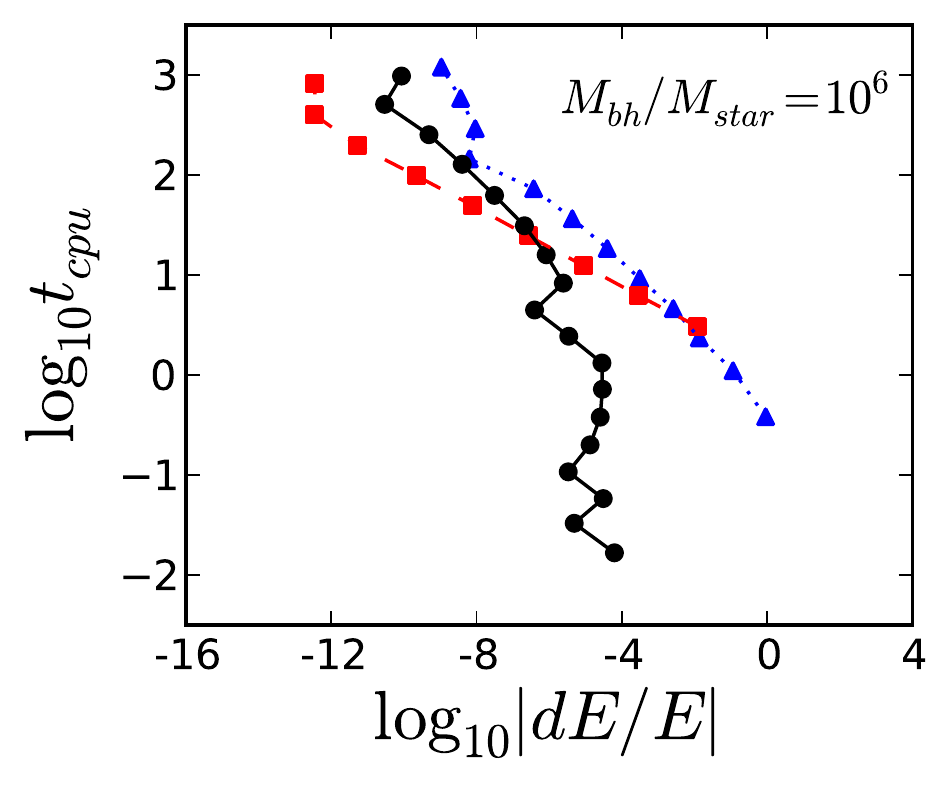}
\includegraphics[scale=0.8]{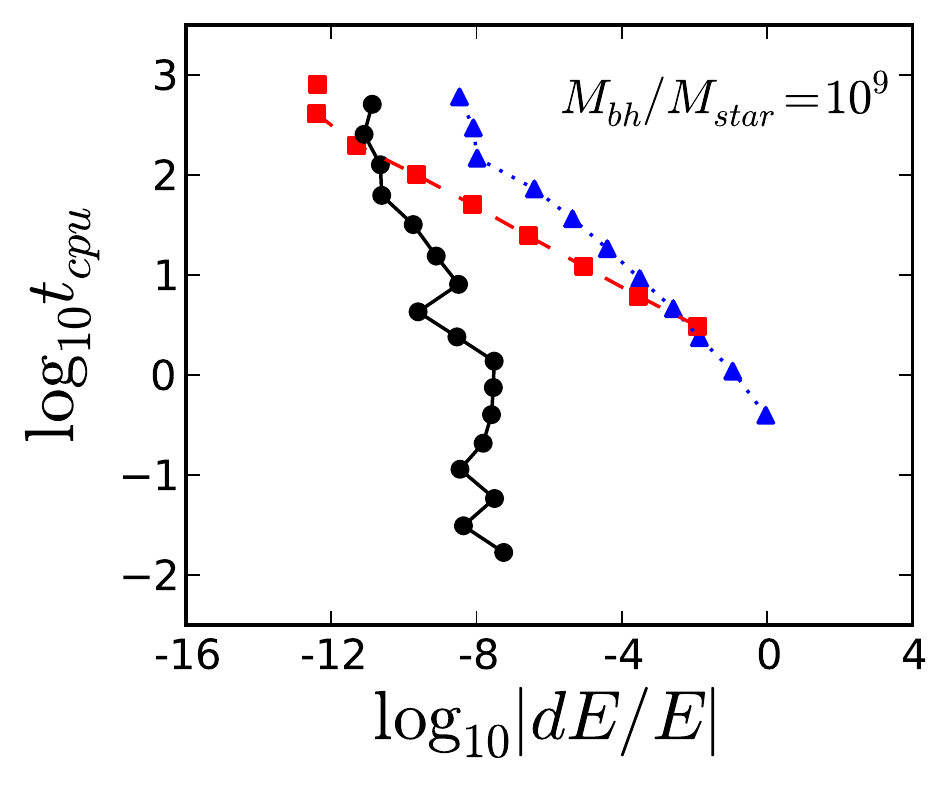}
\includegraphics[scale=0.8]{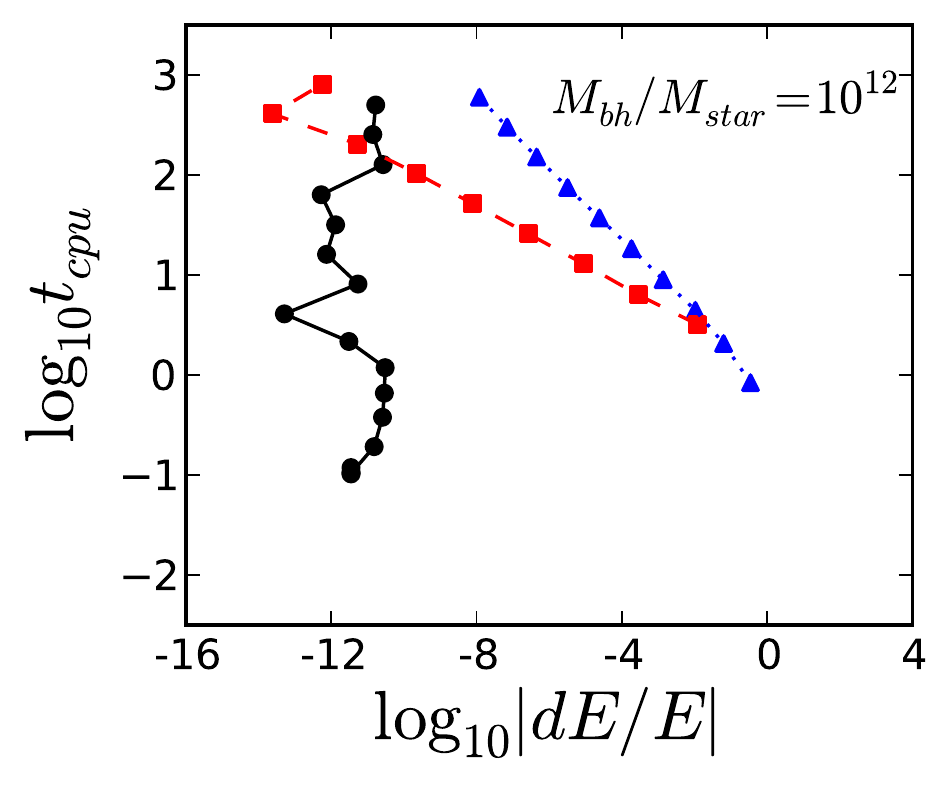}
\caption{For the same system as in Fig.~\ref{fig:03}, the panels show the CPU time (in seconds) vs relative energy error for the following mass ratios: $q\equiv M_{\mathrm{bh}}/M_{\mathrm{star}} = 10^{3}$ (top left), $q=10^{6}$ (top right), $q=10^{9}$ (bottom left) and $q=10^{12}$ (bottom right).}
\label{fig:04}
\end{figure*}

To test how \texttt{Sakura} behaves with a more general $N$-body problem, we use as initial condition a $128$-body Plummer sphere containing a black-hole in its center. We assume equal mass for the stars and construct the system in virial equilibrium but for different black-hole to star mass ratios, $q\equiv M_{\mathrm{bh}}/M_{\mathrm{star}}$, ranging from $q=1$ (no black-hole) to $q=10^{12}$. We performed simulations for each of these initial conditions for $1$ $N$-body time unit. Once again, the performance of \texttt{Sakura} is compared with that of the Leapfrog and standard $4$-th order Hermite integrators. The results are shown in Fig.~\ref{fig:03} which presents the relative energy error as a function of the mass ratio for time-step sizes $\langle \tau \rangle = 10^{-3}, 10^{-4}, 10^{-5}$ (top, middle and bottom lines), and Fig.~\ref{fig:04} which present the CPU time vs relative energy error for different mass ratios: $q = 10^{3}$ (top left), $q=10^{6}$ (top right), $q=10^{9}$ (bottom left) and $q=10^{12}$ (bottom right).

In Fig.~\ref{fig:03} we see that the relative energy error for all three methods initially increases with the mass ratio till the point when $q\sim 10^2$. For larger mass ratios, the behaviour of \texttt{Sakura} clearly differs from the other two methods. While in Leapfrog and Hermite integrators the energy error stabilizes at a certain level, in \texttt{Sakura} we observe a very interesting trend in which its energy error decreases with increasing mass ratio. In other words, \texttt{Sakura} becomes more precise and therefore more efficient when the mass ratio grows, as can be seen in Fig.~\ref{fig:04} for mass ratios (top left), $q=10^{6}$ (top right), $q=10^{9}$ (bottom left) and $q=10^{12}$ (bottom right). An explanation of why these methods behave this way is as follows.

When no dominant massive particle is present in the system ($q\sim 1-10$), after only $1$ $N$-body time unit the system has not evolved for enough time to form a close binary (which is an outcome of strong few-body interactions, see e.g. \citealt{Tanikawa12}). Therefore, in these circumstances most of the particles interact weakly among themselves and all the methods are able to integrate the orbital evolution of stars with relatively good energy conservation. Arround a mass ratio $q\sim 10-10^{3}$ the massive particle quickly forms a binary system with a close neighbour, which eventually experiences several interactions with close perturbers, thus deteriorating the precision of the integration in all three methods. For mass ratios $q\gtrsim 10^{3}$ the orbital motion of stars becomes predominantly Keplerian. In this regime, the orbits in the system become mostly regular, and close encounters between stars become gradually less important. Therefore, the energy error is expected to converge to the truncation error associated to each of these methods. In Leapfrog and Hermite integrators, by decreasing the time-step size the energy conservation is thus improved but it remains approximately at the same level of conservation regardless the mass ratio (for $q\gtrsim 10^{3}$). On the other hand, \texttt{Sakura} departs from a constant level of energy conservation observed in the other two integrators, and becomes increasingly more precise with the mass ratio. This happens because in \texttt{Sakura}, the truncation error comes from two different sources: i) the error due to the \texttt{kepler\_solver}, which is essentially at machine precision, and ii) the error associated to the non-commutativity of $2$-body interactions in close multiple-body encounters. With this knowledge, it is easy to intuitively understand why \texttt{Sakura} becomes more precise with the increase of the mass ratio: simply because the error associated to the non-commutativity of $2$-body interactions becomes less important and, thus the overall error of the integrator converges to that of the \texttt{kepler\_solver}.

For $\langle \tau \rangle \sim 10^{-4}$, which corresponds to the middle lines (for each integrator) in Fig.~\ref{fig:03}, \texttt{Sakura} is $\sim 5$ ($\sim 6$) orders of magnitude more precise than Hermite (Leapfrog), for a mass ratio $q = 10^{6}$. Also, as is shown in Fig.~\ref{fig:04}, \texttt{Sakura}'s performance is similar to Leapfrog, for a mass ratio $q = 10^{3}$, and becomes gradually more efficient than Hermite and Leapfrog, when the mass ratio increases. This happens due to a change in slope of \texttt{Sakura}'s curves in panels showing the CPU time vs relative energy error when the mass ratio goes from $q = 10^{3}$ to $q = 10^{12}$ in Fig.~\ref{fig:04}, which means that for mass ratios $q\gtrsim 10^{12}$, \texttt{Sakura} can give very accurate results ($dE/E \sim 10^{-10} - 10^{-12}$) even when using relatively large time-steps, thus saving a big amount of computational time compared to Leapfrog and Hermite integrators.

\begin{figure}
\centering
\includegraphics[width=\linewidth]{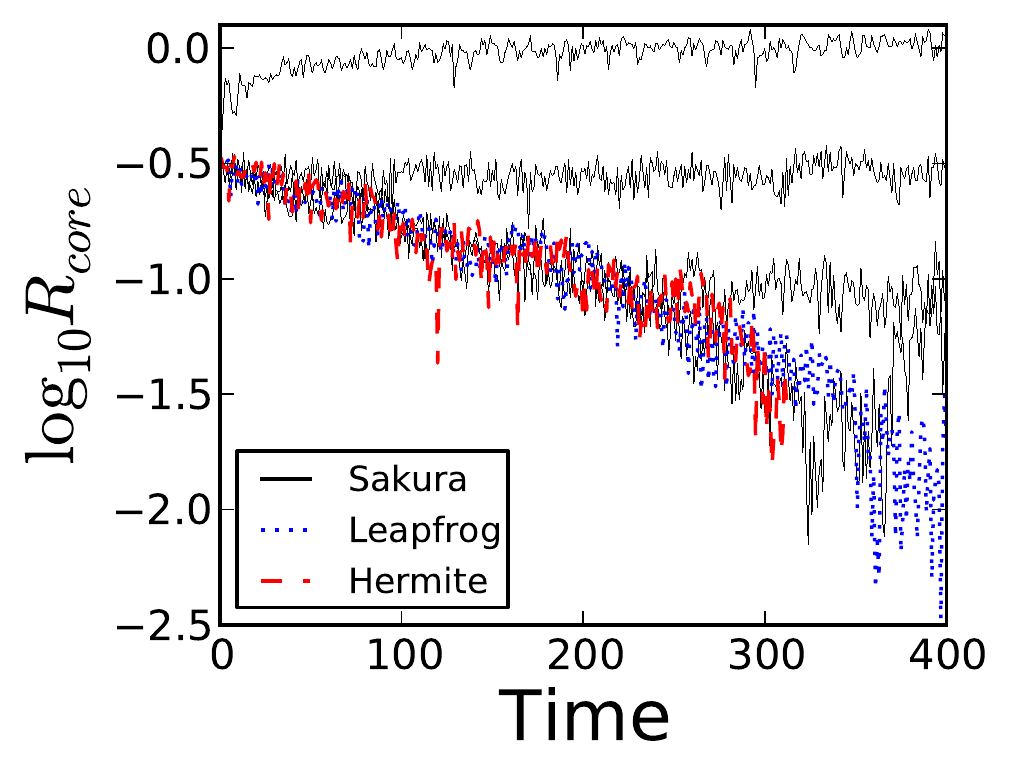}
\caption{Core radius vs simulation time for a 1024-body Plummer sphere. We compare \texttt{Sakura} using different time-step sizes (solid black lines, $\tau = 10^{0}, 10^{-1}, 10^{-2}, 10^{-4}$  from top to bottom) to Leapfrog (dotted blue lines) and standard $4$-th order Hermite (red dashed line), using shared adaptive time-steps with a parameter of precision $\eta \approx 0.03$. All the quantities are presented in $N$-body units.}
\label{fig:05}
\end{figure}

As an additional general $N$-body test we performed a simulation of a $1024$-body system through core collapse using \texttt{Sakura} with several time-step sizes $\tau = 10^{0}, 10^{-1}, 10^{-2}, 10^{-4}$, and the Leapfrog and standard $4$-th order Hermite code using shared adaptive time-steps. For the parameter of precision we choose $\eta = 2^{-5} \approx 0.03$ in order to have a level of energy conservation of about $10^{-4}$ by the moment of core collapse in Hermite integration. In this particular test, we have used a parallel version of \texttt{Sakura} (see section \ref{sec4:parallelization}) running on a 4-core Intel Xeon CPU @2.40 GHz. For the Leapfrog and Hermite codes (which are also parallelized) we setup the number of \texttt{MPI} processes to $4$. In Fig.~\ref{fig:05} we present the time evolution of the core radius using these codes.
We see from this figure that for a sufficiently small time-step size ($\tau\sim 10^{-4}$, lowest black curve in Fig.~\ref{fig:05}) \texttt{Sakura} is able to evolve the system through core collapse. As expected from the exponential orbital instability \citep{Goodman_etal1993}, the results from \texttt{Sakura} slightly differ from Hermite and Leapfrog calculations. Apart from that, the core radius evolution obtained using \texttt{Sakura} follows remarkably well the results from the other two integrators.

In \texttt{Sakura}, the appearance of close binaries does not represent a computational challenge. Therefore, in this simulation no slow down in performance is observed, as is the case in most other $N$-body codes that also try to correctly evolve such compact sub-systems. As a consequence, the most expensive simulation using \texttt{Sakura} (bottom black line in Fig.~\ref{fig:05}) was completed in about three days of CPU time. The Leapfrog integration took about a week of processing time, whereas the Hermite simulation, after more than a month of CPU time (on the same machine), had not been completed, due to the dynamical formation of very close binaries and consequent decrease of the adaptive time-step size.

Although \texttt{Sakura} integrates all pairwise interactions exactly, the presence of close perturbers for a particular $i-j$ pair represents the main source of error during the integration. The reason for that originates from our assumption that each pair of particles can be treated as an independent $2$-body problem during a time-step $\tau$. If $\tau$ is larger than the time scale of interaction between the $i-j$ pair and its perturber, the perturbation will be delayed by $\tau$, leading to spurious integration of a tight multi-component sub-system in an $N$-body simulation. This is a consequence of the non-commutativity of $2$-body interactions. In Fig.~\ref{fig:05}, the use of relatively large time-steps reveals this issue: although the system as a whole stays bound, strong few-body interactions in the cluster core are not correctly integrated and as a consequence the core radius expands. However, by using smaller $\tau$ the numerical issues due to strong perturbations on the $i-j$ pair is diminished and as a consequence \texttt{Sakura} evolves the multi-component sub-systems that may form dynamically during the simulation more precisely. In those calculations, the level of energy conservation at the moment of core collapse stayed within $dE/E\lesssim 10^{-4}$ for Hermite, and $dE/E\lesssim 10^{-2}$ for Leapfrog and \texttt{Sakura} (for the bottom black line in Fig.~\ref{fig:05}), even though \texttt{Sakura} used a constant time-step.

The possibility to include a variable time-step scheme in \texttt{Sakura} might improve its results and is currently under investigation. The fact that \texttt{Sakura} evolves each pair of particles exactly, implies that the time-step criterion does not need to be so restrictive as in the case of traditional integration schemes. For example, if we consider the case of a hierarchical triple system in which the orbital period of the inner binary is a certain factor shorter than the time-scale of interaction between the binary and the outer perturber, we have observed in our tests (not reported here) that choosing a time-step size comparable to the longest time-scale still preserves the binary orbital evolution. In traditional codes, this would not be possible and the inner binary would end up being artificially disrupted if the time-step size has not been decreased to a fraction of its orbital period. Therefore, for \texttt{Sakura} we suspect that a time-step criterion based on the closest perturber distance to a given pair being evolved seems to be a more appropriate choice than an Aarseth-like time-step criterion. We will further discuss this issue on section~\ref{sec5:discussion}.

\section{Parallelization}
\label{sec4:parallelization}

We have implemented three different versions of \texttt{Sakura}: i) a single GPU implementation using \texttt{OpenCL}; ii) a distributed memory parallel implementation using \texttt{MPI}, and iii) a serial implementation in \texttt{C/C++} (used in all the tests presented above, with exception of the one in Fig.~\ref{fig:05}, for which the \texttt{MPI} version was used). The parallelization schemes adopted for distributed memory and GPU versions are quite similar as those adopted for conventional $N$-body codes on those platforms (see \citealt{PortegiesZwart_etal08} and \citealt{NylandHarris2007}, respectively). At the current stage of development our GPU implementation is not yet very efficient due to many branch conditions present in the Kepler-solver.

\begin{figure}
\centering
\includegraphics[width=\linewidth]{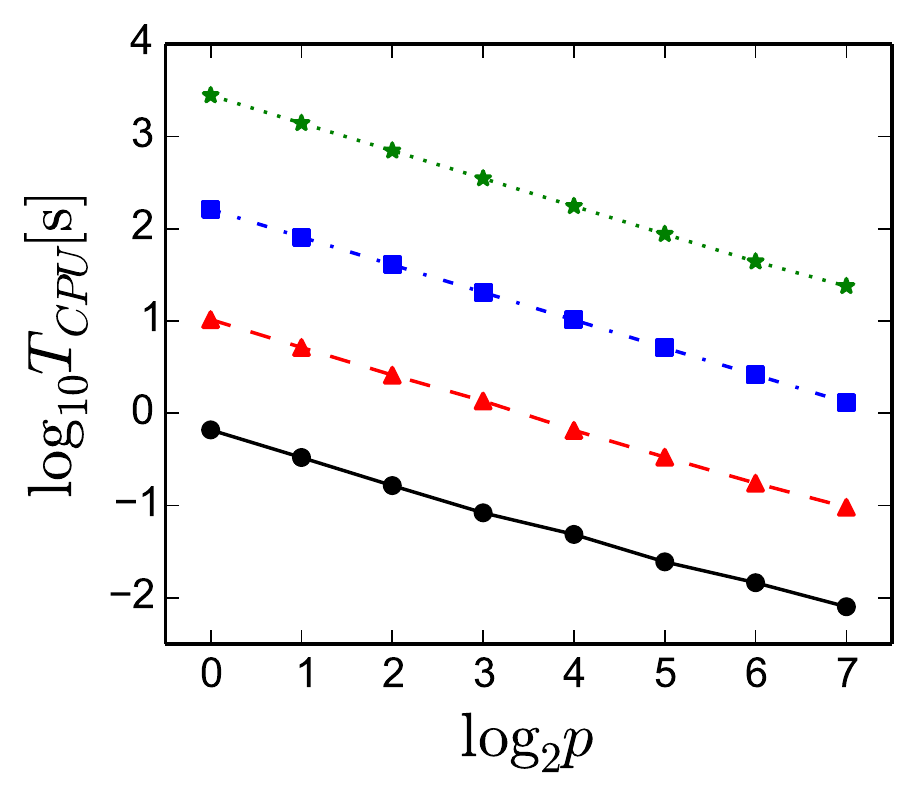}
\includegraphics[width=\linewidth]{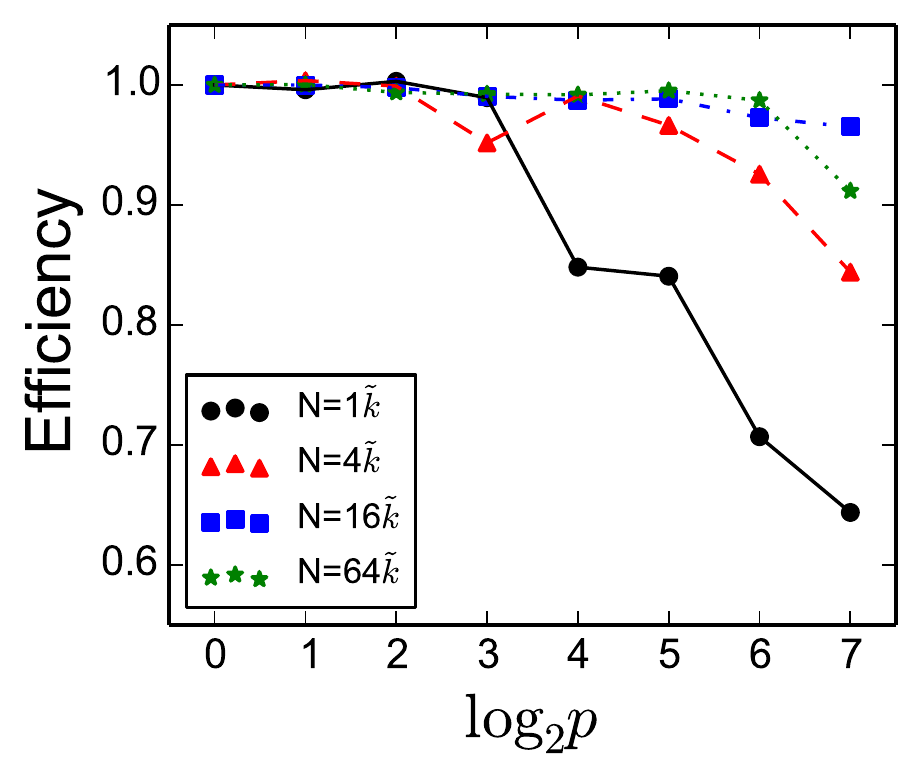}
\caption{For the \texttt{MPI} version of \texttt{Sakura} the plots shows the strong scaling (top panel) and the parallel efficiency (bottom panel) for four different problem sizes: $N = 1\tilde{k}$ (solid black lines), $N = 4\tilde{k}$ (dashed red lines), $N=16\tilde{k}$ (dot-dashed blue lines) and $N=64\tilde{k}$ (dotted green lines). Here, $\tilde{k}$ stands for $1024$ and $p$ is the number of processor cores used for the run.}
\label{fig:06}
\end{figure}

Here we mainly present some performance results using the MPI version of \texttt{Sakura} for tests using up to 128 CPU cores. The test simulations consist of a Plummer sphere with $N$ equal mass particles being integrated for 1 $N$-body time unit. We use four different number of particles $N = 1\tilde{k}$, $4\tilde{k}$, $16\tilde{k}$, $64\tilde{k}$ ($\tilde{k}$ stands for $1024$) and in each case we measure the total wall clock time needed to complete the simulation with different number of cores. In Fig.~\ref{fig:06} we present, for four different problem sizes, the performance measurements in the form of the strong scaling ($T_{CPU}(p)$ vs $p$) and the parallel efficiency:
\begin{eqnarray}
\mathrm{Efficiency} \equiv \frac{T_{CPU}(p)}{p T_{CPU}(1)}\,,\label{eq:29}
\end{eqnarray}
where $T_{CPU}(p)$ is the CPU time measured when using $p$ processor cores.
%Fig.~\ref{fig:06} presents the performance measurements in the form of wall clock time (averaged over 100 steps) and the parallel efficiency obtained for increasing number of cores.

As is evident from the Fig.~\ref{fig:06}, \texttt{Sakura} exhibits an almost perfect strong scaling (top panel) and a remarkably good parallel efficiency (bottom panel). For the worst case scenario presented here ($N = 1\tilde{k}$, using 128 CPU cores), \texttt{Sakura} achieves a parallel efficiency as good as $64\%$, even though the workload in this case is as small as $8$ particles per core. In addition, the strong scaling plot shows that, even in this worst case scenario, the CPU time could still be decreased by using a higher number $p$ of processor cores. For $N > 4\tilde{k}$, the parallel efficiency of \texttt{Sakura} stays very close to $100\%$.

% As is evident from the bottom panel in Fig.~\ref{fig:06}, \texttt{Sakura} exhibits a remarkably good parallel efficiency. For the worst case scenario presented here ($N = 1\tilde{k}$, using 128 CPU cores), \texttt{Sakura} achieves a parallel efficiency as good as $64\%$, even though the workload in this case is as small as $8$ particles per core. For higher $N$, the parallel efficiency of \texttt{Sakura} stays very close to $100\%$.

\section{Summary and Discussion}
\label{sec5:discussion}

We have described a Keplerian-based Hamiltonian splitting for gravitational $N$-body simulations and its implementation in a new code called \texttt{Sakura}. In this method a general $N$-body problem can be solved as a composition of multiple, independent, $2$-body problems. The integration scheme is constructed on the assumption that, during a small time interval $\tau$, each pair of particles in the system can be treated as an independent $2$-body problem. With this splitting an analytical \texttt{kepler\_solver} can be used to accurately, and independently, evolve each $2$-body interaction in the system, thus making the code especially suitable for simulations in which compact primordial binaries or close dynamically formed binaries are present. Hierarchies in which one of the components is a compact binary and systems with a central dominant mass are also examples of physical systems in which \texttt{Sakura} performs well when compared to traditional codes.

Because \texttt{Sakura} can easily handle arbitrarily compact binaries in an $N$-body simulation, the code is able to evolve a star-cluster through core-collapse without much difficulty. In particular, since \texttt{Sakura} can do this even with the use of constant time-steps, the simulation does not suffer from any slow down in performance as is the case in other non-regularized $N$-body codes. As an example, in the $1024$-bodies core-collapse simulation presented in section~\ref{sec3.2:largen}, \texttt{Sakura} was able to complete the run in about $3$ days of CPU time on a $4$-core machine. The same system being integrated with a $4$-th order Hermite integrator took more than one month of CPU time on the same machine, due to a severe slow down in performance after the formation of the first hard binary in the system.

There are, however, some circumstances in which \texttt{Sakura} may not be the most suitable code to perform an $N$-body simulation. For example, for systems in which multiple bodies democratically interact among themselves, \texttt{Sakura} may perform almost as badly as a simple Leapfrog integrator, as demonstrated in the integrations of a figure-eight system in section~\ref{sec3.1:smalln}. This happens because of our underlying assumption that the $N$-body problem can be decomposed in multiple, independent, $2$-body problems. Such decomposition in fact constitutes the main source of error when a given $i-j$ pair is being integrated with a time-step $\tau$ which is larger than the time-scale of the perturbation due to a close neighbour. In many cases this issue may be surpassed by decreasing the constant time-step size used in the simulation. However, the cause of the problem lies on the non-commutativity of $2$-body interactions when multiple bodies are involved in a democratic close encounter. While it is not easy to solve this issue without breaking our Keplerian splitting approach, the introduction of an adaptive time-step scheme in \texttt{Sakura} might alleviate these numerical difficulties and is currently under investigation.

According to some of our tests (not reported in the present paper), a time-step criterion based on the strength of the perturbation on a given $i-j$ pair seems to work relatively well compared to a constant $\tau$. However, this improvement is only significant when close multiple-body encounters take place. On the other hand, one could in principle choose $\tau \sim \min(r_{ij}/v_{ij})$, $\tau \sim \min((r_{ij}/a_{ij})^{1/2})$ or use a traditional Aarseth-like time-step criterion, but we advocate that this may not be the optimal choice because these criteria also include the contribution of the $i-j$ pair itself, which in principle contributes to a severe decrease in time-steps if a close binary is present in the system. In \texttt{Sakura}, these severely short time-steps are not necessary, because the use of a Keplerinan treatment for each pair of particles automatically regularizes every $2$-body interaction in the system. It is only when multiple-body encounters happens that the time-step should adapt itself to properly resolve the approximation of a perturber. Therefore, we stress here our preference for a perturbation-based time-step criterion rather than an Aarseth-like criterion for use in \texttt{Sakura}. Whether or not such perturbation-based criterion is the best choice for \texttt{Sakura} is a matter that will be addressed elsewhere.

Another point we want to emphasize here is the behaviour of \texttt{Sakura} when integrating a system with a central massive black-hole. As shown in Fig.~\ref{fig:03}, the level of energy conservation in Leapfrog and $4$-th order Hermite integrations remains approximately constant with the increase of the black-hole to stellar mass ratio. For \texttt{Sakura}, we found that it performs much better than previous approaches, becoming gradually more precise with the increase of the mass ratio. In particular, for the case of a mass ratio $q = 10^{6}$ \texttt{Sakura} can give $\gtrsim 5$ orders of magnitude better energy conservation than Hermite integrator, being at the same time up to $4$ orders of magnitude faster when the mass ratio increases to $q\gtrsim 10^{9}$. The fact that \texttt{Sakura} can be, at the same time, fast and accurate in this regime, makes this code highly suitable for nearly Keplerian systems where a massive particle dominates the evolution of surrounding particles, such as in planetary systems and galactic nuclei with super-massive black-holes.

Lastly, \texttt{Sakura} has proven to be quite easy to parallelize for distributed memory systems using \texttt{MPI}. The GPU implementation,
even though theoretically easy, is still not totally efficient due to the presence of many branching conditions in the Kepler-solver. In algorithmic terms, the bulk of computation in \texttt{Sakura} occurs inside a double loop, similar to the one used to calculate the acceleration of particles in conventional $N$-body codes. Therefore, we were able to immediately employ existent parallelization schemes in \texttt{Sakura} without much effort. We argue that the fact that our GPU implementation is not yet very efficient is not a problem due to the paralelization scheme itself, but rather due to the poor/inefficient support for branch conditions in current GPUs. A restructure in our Kepler-solver in order to eliminate (or minimize) these branch conditions may address this issue, and possibly speed up even more the \texttt{MPI} version on CPUs, which has already shown a remarkable parallel efficiency, with close to $100\%$ efficiency for $16\tilde{k}$ particles on $128$ cores, and $64\%$ efficiency when using only $8$ particles per core.

% Lastly, \texttt{Sakura} has proven to be quite easy to parallelize for both GPUs and/or distributed memory systems using \texttt{MPI}. In algorithmic terms, the bulk of computation in \texttt{Sakura} occurs inside a double loop, similar to the one used to calculate the acceleration of particles in conventional $N$-body codes. Therefore, we were able to immediately employ existent parallelization schemes in \texttt{Sakura} without much effort. Although our GPU implementation is not yet very efficient due to many branch conditions present in the Kepler-solver, we argue that this issue can be addressed by restructuring the Kepler-solver in order to eliminate (or minimize) these branch conditions, without have to do any modification on the paralelization scheme itself. On the other hand, the \texttt{MPI} version has shown a remarkable parallel efficiency, with close to $100\%$ efficiency for $16\tilde{k}$ particles on $128$ cores, and $64\%$ efficiency when using only $8$ particles per core.

\section*{Acknowledgments}

We thank J. Makino, A. Quillen, I. Pelupessy, M. Fujii, D. Caputo and A. Rimoldi for useful comments that improved the presentation of the paper. The authors are also grateful for fruitful discussion with D.C. Heggie. We also would like to thank the anonymous referee for a critical review, and for suggesting the notation used in section~\ref{sec2:method}. GGF acknowledge the support from CAPES Foundation (Brazil), grant \#5772-11-7. This work was supported by the Netherlands Research Council NWO (grants \#643.200.503, \#639.073.803 and \#614.061.608) and by the Netherlands Research School for Astronomy (NOVA). The computations were performed on the \texttt{Jupiter} and \texttt{LGM} clusters at Leiden Observatory.

\bibliographystyle{mn2e}
\bibliography{refs}

\bsp

\label{lastpage}

\end{document}